\documentclass[11pt, oneside]{article}   	
\usepackage{geometry}                		
\usepackage{graphicx}				
\usepackage{subcaption}
\usepackage{amssymb}
\usepackage{amsmath}

\title{Virus genome sequence classification using features based on nucleotides, words and compression}
\author{T. Wang \hspace{15mm} tingting.wang.11@ucl.ac.uk \\
M. Herbster \hspace{15mm} m.herbster@cs.ucl.ac.uk \\
I.S. Mian \hspace{28mm} s.mian@cs.ucl.ac.uk \\
Department of Computer Science, University College London \\
66 Gower Street, London WC1E 6BT, UK}

\date{}							

\begin{document}
\maketitle

\noindent \textbf{Abstract} The International Committee on Taxonomy of Viruses (ICTV) develops, refines and maintains a universal virus taxonomy; Order is the highest taxon in the branching hierarchy of recognised viral taxa. Historically, ICTV (sub)committees have classified viruses on the basis of morphological characteristics and various other attributes. Today, virtually all new viral genomes are assembled from metagenomic datasets and are not linked directly to biological agents. Thus, placing a virus into a taxonomic scheme solely from primary genome structure is an increasingly important problem. Various simple descriptive statistics of a viral genome sequence have been used successfully for virus classification. Here, we use the NCBI's viral and viroid reference sequence collection (RefSeq) and a common experimental framework to compare the performance of different genome sequence-derived features and classifiers in the task of assigning a virus to one of seven ICTV Orders. The nucleotide-, word-, and compression-based features we consider include genome length, the $k$-mer Natural Vector ($k = 1, \cdots, 6$) and its derivatives, return time distribution, and general-purpose and DNA-specific compression ratios; the classifiers used are the k Nearest Neighbours method (k-NN) and the Support Vector Machine (SVM) with a radial basis function kernel. The combination of genome length and k-NN has the worst, yet still respectable, performance (mean error rate of 0.137); the best performance is achieved using 4-mer counts and SVM (mean error rate of 0.006). We investigate the main causes of misclassification, explore which viruses are more difficult to classify, and use the best performing combination to predict the Orders of 1,834 unclassified viruses. A subsequent version of RefSeq assigned Orders to 17 of these previously unlabelled viruses. Since 16 of our predictions match these assignments, our approach could aid virologists dealing with viruses that are known only from sequence data.


\section{Introduction}
Virus classification is the task of placing a virus into a taxonomic scheme, a system that groups organisms together on the basis of shared features such as evolutionary history, genotypic and phenotypic characteristics, and biological properties \cite{King2012}. For researchers and clinicians, the taxonomy provides insights into virus biology \cite{KrupovicJV2010,King2012,SimmondsJGV2015}, enables the development of reliable diagnostic procedures, enhances epidemiological studies and facilitates the treatment and prevention of disease \cite{HyypiJGV1997}.

The International Committee on Taxonomy of Viruses (ICTV) is the authority charged with classifying viruses into taxa and naming them \cite{King2012}. It defines a hierarchy of taxa including Order (most general), Family, Subfamily, Genus and Species (most specific). ICTV (sub)committees use verifiable genotype- and phenotype-related data to propose the (re)assignment of a virus to a taxon. This is a labour-intensive and time-consuming approach, which is ill-suited to handle the influx of metagenomic sequence data produced by modern sequencing technologies. For this reason, the ICTV Executive Committee has proposed new pathways that allow viruses identified from only metagenomics data to be admitted into the current formal taxonomy \cite{SimmondsNRM2017}. However, although utilising only primary genome structure to classify viruses can be effective and efficient \cite{VingaB2003,SchwendeBB2014,SongBB2014}, it may not always yield significant biological insight \cite{SimmondsJGV2015,KrupovicJV2010}.

Computational methods for comparing virus genome sequences and thus classifying viruses fall into two categories according to how the distance between sequences is defined. Alignment-based methods align sequences and then use an alignment score (local or global, multiple or pairwise) to determine a distance \cite{BaoAV2014,MuhirePO2014,KatohNAR2002,EdgarNAR2004,LarkinB2007}. Such methods do not perform well on highly diverged viruses whose sequences cannot be aligned reliably, and are hard to scale to massive datasets due to issues of computational complexity \cite{WangJCB1994}. In contrast, alignment-free methods map sequences to points in a feature space and use the distance between points in that space as the distance between sequences. Sequence representations and analytical approaches include descriptive statistics such as nucleobase occurrence and positional information \cite{WenJTB2014,YuDR2010,DengPO2011,WenG2014}, matrix invariants \cite{HeIEJMD2002}, genome signal processing \cite{YinJTB2015}, curves \cite{ZhangCG2014}, and images \cite{AlmeidaBB2014}. The complementary nature of alignment-based and alignment-free methods has been exploited to classify long and short sequence reads from next-generation sequencing technology \cite{BorozanB2015}.

This work focuses on three categories of alignment-free transformations of long complete genome sequences of varying lengths into much shorter fixed length vectors of simple yet intuitive features calculated directly from a single sequence; below we list some specific feature vector representations and their application to questions related to the phylogeny, taxonomy and genome biology of not just viruses but also cellular organisms.

{\em Nucleotide-based features}: summary statistics of individual nucleotides. Genome length, the number of nucleotides in a sequence, has been shown to distinguish a number of different organisms~\cite{LiPO2014}. The Natural Vector (NV)~\cite{DengPO2011} comprises the count of each nucleobase, together with the mean and normalised variance of that nucleobase's position in the sequence. It has been used to predict the ICTV Order of both non-segmented viruses (whose genetic material exists as a single nucleic acid molecule) \cite{YuPO2013} and multi-segmented viruses (whose genetic material exists as two or more molecules) \cite{HuangMPE2014}.

{\em Word-based features}: summary statistics of subsequences of $k$ consecutive nucleotides ($k$-mers). The $k$-mer NV~\cite{WenG2014} comprises the count of each $k$-mer together with the mean and normalised variance of its position. It has been utilised to determine phylogenetic relationships between mammalian mitochondrial genome sequences and 18S ribosomal RNA sequences without the need for a model of evolution. An ensemble distance measure of $k$-mer count vectors and NV has been used for phylogenetic analysis of multi-segmented viruses \cite{HuangJTB2016}. The Generalized Vector, a concatenation of the $k$-mer NV and the Complete Composite Vector, has been employed for virus classification \cite{HernandezACUL2013}. The Return Time Distribution (RTD)~\cite{KolekarACP2010,KolekarMPE2012,KumarIS2016} describes the mean and variance of the successive appearances of $k$-mers. It has been used for phylogenetic analysis of the ICTV Family \textit{Flaviviridae} and subtyping of \textit{Dengue} viruses \cite{KolekarMPE2012} as well as genotyping of \textit{Mumps} viruses \cite{KolekarIR2011}.

{\em Compression-based features}: the compressibility of sequences when treated as ``natural language'' text or nucleic acid strings. Both general-purpose algorithms (for example, gzip, bzip2, xz and zip) and algorithms tailored to DNA sequences (for example, DELIMINATE, MFCompress, LEON and Nucleotide-Pair Encoding) have been used for storage, retrieval, and transmission of high throughput DNA sequencing data~\cite{BakrAJBR2013,PratasThesis2016,SardarazJBCB2016,HosseiniI2016,asgari2018nucleotide}.

Here, we employ a uniform experimental framework to investigate the performance of different alignment-free representations of complete viral genome sequences (vectors constructed from nucleotide-, word- and compression-based features) and different classification methods (k-NN and SVM) on the same virus classification problem (assigning one of seven ICTV Orders to a sequence).
The paper is structured as follows.

Section~\ref{sec_feature} reviews the feature vectors analysed: genome length, the $k$-mer NV and its derivatives, RTD, and compression ratios. We consider $k$-mers based on both the standard ACGT alphabet and three 2-letter alternatives. Section~\ref{sec_classf} summarises the two classifiers employed: k-NN and SVM with a radial basis function kernel. Section~\ref{sec_exp} describes the dataset: the viral and viroid reference sequence (RefSeq) collection from the National Center for Biotechnology Information (NCBI).

Sections~\ref{sec_visualisation}~and~\ref{sec_results} discuss our exploratory data analysis (EDA) and classification experiments. We visualise the relationship between various underlying statistical properties of complete viral genome sequences and the ICTV Orders assigned to viruses. We determine the error rates of different combinations of feature vectors and classifiers in the task of predicting Order. We ascertain the components of a feature vector important for classification, and the main causes of misclassification. We utilise the best performing feature vector-classifier combination (AGCT 4-mer counts plus SVM) to identify sequences that are difficult to classify and to predict the Order for sequences without this assignment. Section~\ref{sec_discussion} explores the implications of our findings. We discuss the problem of imbalanced classes and the validity of our Order predictions in light of the updated RefSeq dataset.

Section~\ref{sec_future} suggests directions for future research. We suggest hierarchical classification as a strategy for assigning not only Order but all taxa in the ICTV hierarchy. We discuss distributed representations of biological sequences, a methodology proposed originally in Natural Language Processing to identify terms with a similar linguistic context. We propose graph-based semi-supervised learning as a way to predict the phenotypic properties of viruses and to infer phylogenetic relationships between them from the large and rapidly increasing corpus of viral genome sequences and the small collection of knowledge about the biological properties of virions.

\section{Sequences as vectors of features}
\label{sec_feature}

\subsection{Nucleotide-based features}

\subsubsection{Genome Length}

Length, the total number of nucleotides, is the simplest descriptive statistic of a genome sequence \cite{LiPO2014}. 

\subsubsection{Natural Vector (NV) and its derivatives}

The NV methodology~\cite{DengPO2011} recasts a variable length nucleic acid sequence as a fixed-length low dimensional vector of global composition- and location-based statistics for each nucleobase present in the sequence. The simple and intuitive statistics used for this mapping are the count of each nucleobase together with the mean and variance of its position in the sequence. Let $S = s_1 s_2 \cdots s_n$ be a sequence of length $n$ where $s_i, i = 1, 2, \cdots, n$ is a letter in the nucleobase alphabet, typically $s \in \{A, C, G, T\}$. For a nucleobase $s$, the count $n_s$, mean position in the sequence $\mu_s$ and the normalised variance of the position $d_s$ are defined as
\begin{align}
n_s    &= \sum_{i = 1}^{n} I(s = s_i) \nonumber \\
\mu_s  &= \frac{1}{n_s} \sum_{i = 1}^{n} i \ I(s = s_i) \nonumber \\
d_s  &= \frac{1}{n_s n} \sum_{i = 1}^{n} (i - \mu_s)^2 I(s = s_i) \nonumber
\end{align}
where $I(s = s_i)$ is the indicator function which equals 1 when $s = s_i$ and 0 otherwise.

The 12-dimensional NV representation of a nucleic acid sequence $S$ used in earlier work \cite{YuPO2013} contains the following variables,
\begin{equation}
NV12 = [\textbf{n}, \boldsymbol{\mu}, \textbf{d}] \nonumber
\end{equation}
where $\textbf{n}$, $\boldsymbol{\mu}$ and $\textbf{d}$ are the following vectors  
\begin{align}
\textbf{n} &= [n_A, n_C, n_G, n_T] \nonumber \\
\boldsymbol{\mu}  &= [\mu_A, \mu_C, \mu_G, \mu_T] \nonumber \\
\textbf{d} &= [d_A, d_C, d_G, d_T] \nonumber
\end{align}

We extend this previous work by investigating the contribution of components of NV12.  We derive two ``simpler'' feature vectors: NV4 (counts) and NV8 (counts and mean positions).

\subsubsection{Summary of nucleotide-based features}

Table \ref{features_nuc} lists the names, dimensions and components of the nucleotide-based feature vector representations of the viral genome sequences we studied.

\begin{table}[!ht]
\caption{Feature vectors based on nucleotide statistics.}
\label{features_nuc}
\centering
\begin{tabular}{l l l} 
\hline
Name    & Dimension  & Variable(s) \\ 
\hline
Length  &  1         & $n$\\ 
NV4     &  4         & $n_A, n_C, n_G, n_T$\\ 
NV8     &  8         & $n_A, n_C, n_G, n_T$, \\
        &            & $\mu_A, \mu_C, \mu_G, \mu_T$\\ 
NV12    & 12         & $n_A, n_C, n_G, n_T$, \\
        &            & $\mu_A, \mu_C, \mu_G, \mu_T$, \\
        &            & $d_A, d_C, d_G, d_T$\\ \hline
\end{tabular}
\end{table}

\subsection{Word-based features}

A word, or $k$-mer, is a short sequence of $k$ consecutive nucleotides $w = s_i s_{i+1} \cdots s_{i+k-1}$. The $k$-mers in a genome sequence can be extracted by sliding a window of length $k$ over the sequence from position $s_1$ to position $s_{(n-k+1)}$.

\subsubsection{$k$-mer NV and its derivatives}

$k$-mers and the NV model have been combined \cite{WenG2014} by generalising $\textbf{n}$, $\boldsymbol{\mu}$ and $\textbf{d}$ to the counts, mean positions and normalised variances of all possible $k$-mers,
\begin{align}
\textbf{n} &= [n_{w_1}, n_{w_2}, \cdots, n_{w_{a^k}}] \nonumber \\
\boldsymbol{\mu}  &= [\mu_{w_1}, \mu_{w_2}, \cdots, \mu_{w_{a^k}}] \nonumber \\
\textbf{d} &= [d_{w_1}, d_{w_2}, \cdots, d_{w_{a^k}}] \nonumber
\end{align}
For a nucleobase alphabet of size $a$, there are $a^k$ possible $k$-mers so the dimensionality of the $k$-mer NV (total number of features) is $3 \times a^k$; the count, mean position and normalised variance of $k$-mers that are absent from a sequence are set to zero. The 1-mer NV is equivalent to NV12 when the alphabet is $\{A, C, G, T\}$.

\subsubsection{Return Time Distribution (RTD)}


The RTD \cite{KolekarACP2010,KolekarMPE2012,KumarIS2016} takes into account both the frequency and relative order of $k$-mers in a sequence. The return time is the number of nucleotides between successive appearances of a particular $k$-mer and the RTD is the frequency distribution of those times. The RTD feature vector consists of the means and variances of the RTDs of all $k$-mers.

Let $p_{w_i}^t$ denote the position of the $t$-th occurrence of $k$-mer $w_i$ in a sequence. If the $k$-mer occurs $T$ times in total, then the $t$-th return time is 
\begin{equation}
\Delta p_{w_i}^t = p_{w_i}^t - p_{w_i}^{t-1},\quad (2 \leq t \leq T).  \nonumber
\end{equation}
The RTD is summarised by the mean and variance of the series of return times,
\begin{align}
\mu_{\Delta p_{w_i}^t}  &= \frac{1}{T-1} \sum_{t = 2}^{T} \Delta p_{w_i}^t,\quad (T \geq 2), \nonumber \\
  d_{\Delta p_{w_i}^t}  &= \frac{1}{T-1} \sum_{t = 2}^{T} (\Delta p_{w_i}^t - \mu_{\Delta p_{w_i}^t})^2,\quad (T \geq 3). \nonumber
\end{align}

\subsubsection{Nucleobase alphabets}

We investigated the standard 4-letter alphabet and three reduced 2-letter nucleotide alphabets (Table~\ref{alphabets}).

\begin{table}[!ht]
\caption{4- and 2-letter alphabets used to construct feature vectors.}
\label{alphabets}
\centering
\begin{tabular}{l l l l l} \hline
Alphabet & Size   & Symbol & Name & Nucleobases \\ \hline
ACGT     & 4      & A      & Adenine & A \\ 
         &        & C      & Cytosine & C \\ 
	 &        & G      & Guanine & G \\ 
	 &        & T      & Thymine & T \\ 
SW       & 2      & S      & Strong & C, G \\ 
	 &        & W      & Weak & A, T \\ 
RY       & 2      & R      & puRine & A, G \\ 
	 &        & Y      & pYrimidine & C, T \\ 
MK       & 2      & M      & aMino & A, C \\ 
	 &        & K      & Keto & G, T \\ \hline
\end{tabular}
\end{table}

\subsubsection{Summary of word-based features}

Table~\ref{features_kmer} lists the features, alphabets and dimensions of the word-based feature vector representation of the viral genome sequences we studied. ``Concat-$k$-mer NV'' is a concatenation of the NV of $k$-mers and all their subwords; NV12 is equivalent to the $k$-mer NV when $k = 1$ and the alphabet is ACGT.

\begin{table}[!ht]
\caption{Feature vectors based on $k$-mer statistics ($k = 1, \cdots, 6$) and different alphabets.}
\label{features_kmer}
\centering
\begin{tabular}{l l l l l l l l} \hline
Variables   & Alphabet & \multicolumn{6}{c}{Dimension of $k$-mer feature vector}\\ \cline{3-8}
            &  & $1$ & $2$ & $3$ & $4$ & $5$ & $6$ \\ \hline
Counts      & SW & 2 & 4 & 8 & 16 & 32 & 64 \\ 
Counts      & RY & 2 & 4 & 8 & 16 & 32 & 64 \\ 
Counts      & MK & 2 & 4 & 8 & 16 & 32 & 64 \\ 
Concat-     & SW,  & 6 & 12 & 24 & 48 & 96 & 192 \\ 
counts      & RY, MK &&&&&& \\ 
Counts      & ACGT & 4 & 16 & 64 & 256 & 1,024 & 4,096 \\ 
RTD         & ACGT & 8 & 32 & 128 & 512 & 2,048 & 8,192 \\ 
Counts, RTD & ACGT & 12 & 48 & 192 & 768 & 3,072 & 12,288 \\ 
$k$-mer NV  & ACGT & 12 & 48 & 192 & 768 & 3,072 & 12,288 \\ 
Concat-     & ACGT & 12 & 60 & 252 & 1,020 & 4,092 & 16,380 \\ 
$k$-mer NV  & & &&&&& \\ \hline
\end{tabular}
\end{table}

\subsection{Compression-based features}

Algorithmic information theory has been used to construct a distance metric between (mitochondrial) genome sequences \cite{LiB2001}. Kolmogorov complexity \cite{Kolmogorov1963IJS} is a measure of the computational resources needed to specify an object. It reflects the complexity of the underlying structure of the object and is defined as the length of the shortest computer program (in a predetermined programming language) that produces the object. There are no absolute measures of Kolmogorov complexity so it is estimated using (efficient) compression algorithms \cite{VyuginCJ1999}. The compression ratio, defined as the ratio between the compressed and original file size, reflects the ``complexity'' of the content of the file. We construct feature vectors for a viral genome sequence by concatenating compression ratios obtained using four general-purpose compression tools and three DNA-specific compression tools that are specialised for FASTA files, the format of our dataset \cite{BakrAJBR2013,PratasThesis2016,SardarazJBCB2016,HosseiniI2016}.

\subsubsection{General-purpose tools for compressing text}

{\em Gzip} uses Lempel-Ziv coding (LZ77). The Lempel-Ziv algorithm \cite{ZivTIT1977} is a dictionary compression scheme that works by finding duplicated strings in the input text and replacing those duplicates with a reference (distance and length) to the original string. The distances are compressed with one Huffman tree, and the lengths with another. The gzip implementation limits distances to 32K bytes and lengths to 258 bytes. 

{\em Bzip2} uses the Burrows-Wheeler (BW) block sorting text compression algorithm \cite{Burrows1994} along with Huffman coding \cite{HuffmanIRE1952}. The BW algorithm applies a reversible transformation to a block of input text with the aim of bringing the same characters closer together, making the transformed text easier to compress with simple algorithms.

{\em Xz} uses the Lempel-Ziv-Markov chain (LZMA) compression algorithm \cite{LZMA}. LZMA uses a variant of LZ77 with large dictionary sizes and support for repeatedly used match distances, with the output encoded with a range encoder. Range encoding is an entropy coding method which is similar to arithmetic coding. It differs in that digits can be expressed in any base, rather than just base 2 (as bits). 

{\em Zip} supports a number of compression algorithms. By default (and in our usage), this Unix/Linux file compression and packaging utility uses the DEFLATE compression algorithm, which itself uses a combination of LZ77 and Huffman coding \cite{Deutsch1996}.

\subsubsection{DNA-specific tools for compressing sequences}

{\em DELIMINATE}\cite{MohammedB2012} (Delta encoding and progressive ELIMINATion of nucleotide characters) is a fast and efficient method for lossless compression of genomic sequences. During compression, the positions of the two most frequently occurring bases are delta encoded to be stored as differences rather than absolute values. These bases are subsequently eliminated from the sequence, and the remaining bases are then represented with a binary code. This method outperforms general-purpose tools and the compression gains on large sequence datasets are dramatically higher.

{\em MFCompress} \cite{PinhoB2013} (Multi-fasta and Fasta Compression) is a lossless compression method that is claimed to provide better compression than DELIMINATE for similar compression and decompression times. It relies on multiple competing finite-context models and arithmetic coding \cite{Pinho2011PO}. One finite-context model is used for encoding the header and multiple models are employed for the main stream of the DNA sequences.

{\em LEON} \cite{BenoitBB2015} is a method for compression of data generated from high throughput sequencing techniques. It is based on a reference probabilistic de Bruijn Graph, built from the set of short sequences (called reads) acquired by a sequencing device and stored in a Bloom filter \cite{BloomCACM1970}, a space-efficient probabilistic data structure used to test whether an element is a member of a set. Each read is encoded as a path in this graph, by memorising an anchoring $k$-mer and a list of bifurcations, enough to rebuild it from the graph.

\subsubsection{Summary of compression-based feature vectors}

Table~\ref{feature_compRatio} lists the names, tools and features, dimensions and variables of the compression-based feature vector representations of viral genome sequences we studied. $r_{tool}$ is the compression ratio computed using $tool$.

\begin{table}[!ht]
\caption{Feature vectors based on compression ratios calculated using different compression tools.}
\label{feature_compRatio}
\centering
\begin{tabular}{l l l l} \hline
Name  & Tool(s), Feature & Dimension & Variable(s) \\ \hline
CRB   & bzip2 & 1 & $r_{bzip2}$ \\ 
CRL   & LEON  & 1 & $r_{LEON}$ \\ 
CRBL  & bzip2,     Length & 2 & $r_{bzip2}, log(Length)$ \\ 
CRLL  & LEON,     Length & 2 & $r_{LEON}, log(Length)$ \\ 
CRLB  & LEON, bzip2 & 2 & $r_{LEON}, r_{bzip2}$ \\ 
CRDNA & DNA-specific & 3 & $r_{DELIMINATE}$, \\
&&& $r_{MFCompress}, r_{LEON}$ \\
CRGP  & General-purpose & 4 & $r_{bzip2}, r_{gip}, r_{xz}, r_{zip}$ \\ 
CRA   & All & 7 & $r_{bzip2}, r_{gip}, r_{xz}, r_{zip}$  \\ 
      &     & & $r_{DELIMINATE}$, \\
&&& $r_{MFCompress}, r_{LEON}$ \\ 
\hline
\end{tabular}
\end{table}

\section{Classifiers}
\label{sec_classf}

\subsection{k-Nearest Neighbours (k-NN)} 

The k-NN \cite{AltmanAS1992} classifier was used in previous studies of virus classification \cite{YuPO2013,HernandezACUL2013}. To predict the label of a new virus genome sequence, the method first computes the distance between the feature vector of the given sequence and that of every other sequence in the training set. A prediction is made using the majority vote of labels of its k nearest neighbours (closest sequences) where k is a parameter of the model.

We used the {\tt knn} function implemented in the R package {\tt class} \cite{Rclass}.

\subsection{Support Vector Machine (SVM)}

The SVM~\cite{CortesML1995,CristianiniSVMKM2000,TaylorKMPA2004} is a binary classifier that aims to find a separator that best separates samples from different classes. The optimisation goal is to find a hyperplane that maximises the margin, the distance of the samples closest to the hyperplane. Given training samples $(\textbf{x}_i, y_i), i = 1, \cdots, m$, where $\textbf{x}_i \in R^d$ is a feature vector of dimension $d$ and $y_i \in \{-1, 1\}$ is the label of the $i$-th sample, the primal form of the classical soft-margin problem is formulated as
\begin{align} 
\min_{\textbf{w}, b} & \indent \frac{1}{2} ||\textbf{w}||_2^2 + C \sum_{i=1}^m \epsilon_i \nonumber \\
s.t. & \indent y_i (\textbf{w}^T \textbf{x}_i + b) \geq 1 - \epsilon_i \nonumber \\
     & \indent \epsilon_i \geq 0, i = 1, \cdots, m. \nonumber
\end{align}
The slack variables $\epsilon_i (i=1,\ldots,m)$ ``soften'' inversely with respect to $C$ the optimisation problem of finding the maximum margin hyperplane thereby accommodating incorrectly labelled (noisy) data points \cite{Vapnik1998}. The dual form of the optimisation problem with a kernel function is 
\begin{align}
\max_{\alpha_i} & \indent -\frac{1}{2} \sum_{i,j=1}^m \alpha_i \alpha_j y_i y_j K(\textbf{x}_i, \textbf{x}_j) + \sum_i \alpha_i \nonumber \\
s.t. & \indent \sum_i y_i \alpha_i = 0 \nonumber \\
	 & \indent 0 \leq \alpha_i \leq C, i = 1, \cdots, m. \nonumber
\end{align}
The kernel function $K(\textbf{x}_i, \textbf{x}_j) =\: <\phi(\textbf{x}_i), \phi(\textbf{x}_j)>$ is a measure of similarity between two transformed data points, $\phi(\textbf{x}_i)$ and $\phi(\textbf{x}_j)$, where $\phi(.)$ is a feature map that projects the original feature vector $\textbf{x}_i$ into a higher dimensional space in order to enhance the separability of the data points. One effective kernel function is the radial basis function (RBF) $K({\bf x}, {\bf y}) = \exp(-\gamma||\textbf{x}_i - \textbf{x}_j||^2)$, where $\gamma$ is the inverse ``width'' of the kernel \cite{VertKMCB2004}. A new sample $\textbf{x}$ is classified as $\mbox{sign}(\sum_{i=1}^m {\alpha}_i y_i K(\textbf{x}_i, \textbf{x}) + b)$, where ${\alpha}_i$ are the optimised parameters and ${b} = y_j - \sum_{i=1}^m {\alpha_i} y_i K(\textbf{x}_i, \textbf{x}_j)$ for any $j$ such that $0<\alpha_j<C$.

We used the {\tt svm} function implemented in the R package {\tt e1071} \cite{Re1071}, which is an interface to {\tt libsvm} \cite{ChangTIST2011}. The RBF-SVM has two parameters, the kernel width $\gamma$ and the regularisation constant $C$. For multi-class problems, it adopts a ``one-vs-one'' approach where the label of a test data point is determined by a majority vote of all the pairwise SVMs.

\section{Dataset}
\label{sec_exp}

The ICTV defines a hierarchy of taxa encompassing Order (most general), Family, Subfamily, Genus and Species (most specific). We retrieved the entire viral and viroid RefSeq collection from the ``Viruses'' directory of the NCBI FTP site on 18$^{th}$ September 2015 \cite{ftpNCBIVirusGenome}. Each RefSeq file contains complete genome sequence data and annotation data for a virus species; any ICTV taxa assigned to the virus are present in the ``ORGANISM'' field of the annotations. The 4,420 RefSeq files downloaded span 3,910 non-segmented viruses $-$ of which 211 are satellites $-$ and 510 multi-segmented viruses. We analysed the complete genome sequences of the 3,699 non-satellite non-segmented viruses and considered only the ICTV Order assigned to a virus. Table~\ref{virusInSchemes} provides details regarding the dataset we studied.

\begin{table}[!ht]
\caption{Non-satellite non-segmented viruses studied.}
\label{virusInSchemes}
The name and our abbreviation for an ICTV Order plus the number of complete virus genome sequences with that taxonomic label are shown. Labelled: the number of sequences with an Order label. Unlabelled: the number of sequences without an Order label. Total: the number of labelled and unlabelled sequences. 
\begin{tabular}{l l l} \hline
ICTV Order & Abbreviation & Number of sequences \\ 
\hline
Caudovirales       & C & 1,281 \\ 
Herpesvirales      & H & 66   \\ 
Ligamenvirales    & L & 13   \\ 
Mononegavirales & M & 156 \\ 
Nidovirales          & N & 67 \\ 
Picornavirales     & P & 135  \\ 
Tymovirales        & T & 147  \\ 
\hline
\bf Labelled            & & 1,865  \\ 
Unlabelled             & & 1,834  \\ 
\hline
\bf Total                  & & 3,699  \\ 
\hline
\end{tabular}
\end{table}

In 2016 and 2017, the ICTV added two new Orders, {\em Bunyavirales} and {\em Ortervirales}, yielding a current total of 9 named Orders. Since ratification occurred after the research described in this work was performed, we consider only the seven Orders defined in 2015. 

\section{Exploratory data analysis}
\label{sec_visualisation}

We use a variety of techniques to visualise different statistical properties of the complete virus genome sequences stratified by ICTV Order. The intuition and qualitative understanding gained by uncovering features which contribute to the separability of taxonomic classes will assist us in formulating hypotheses and interpreting the results of the classification experiments described later.

\subsection{Methods}

\subsubsection{Box plot}

This non-parametric visualisation is a convenient way to show key properties of a dataset: the bottom, internal and top bands of the box are the first, second (median) and third quartiles with their spacing indicating the degree of dispersion and skewness; the whiskers are the minimum and maximum values of inliers; and the circles are outliers.

We used the {\tt boxplot} function implemented in the R package {\tt graphics} \cite{Rgraphics} to visualise the distributions of the log transformed values of various nucleotide-based features.

\subsubsection{t-distributed Stochastic Neighbour Embedding (t-SNE) plot}

This is a visualisation of the output of an unsupervised non-linear dimensionality reduction technique that projects high-dimensional data into a low-dimensional space. Similar (dissimilar) objects are modelled as nearby (distant) points.

We used the {\tt Rtsne} function implemented in the R package {\tt Rtsne} \cite{Rtsne} to visualise various feature vectors in a 2-dimension space. Using a random seed and feature vectors for all labelled and unlabelled sequences, we compute the coordinates for every virus in our dataset. We display the resultant embedding as a scatter plot with points coloured and displayed according to the following schemes. Colour can indicate the taxonomic class assigned to a virus: grey represents unlabelled viruses; red, yellow, brown, green, light blue, dark blue, and pink indicate ICTV Order C, H, L, M, N, P, and T respectively (Table~\ref{virusInSchemes}). Point style indicates the taxonomic class assigned to a virus: circles represent unlabelled viruses; the one-letter abbreviations indicate ICTV Order. Colour can also indicate the normalised length of a sequence: normalisation is performed by scaling the natural log transformed Length to the range between 0 and 1, corresponding to the colours red and blue.

\subsubsection{Ternary (simplex) plot or de Finetti diagram}

This visualisation is a concise way to show the relationship between three variables in a two-dimensional space. The coordinate system is organised as an equilateral triangle whose vertices are associated with three variables; the distance from a point to a vertex is inversely proportional to the magnitude of the value of the variable associated with the vertex.

We used the {\tt ggtern} function implemented in the R package {\tt ggtern} \cite{Rggtern} to visualise the composition of sequences using three 2-letter alphabets. The precise location of a point depends on the physicochemical property used to group the four nucleobases: strong/S (G, C) or weak/W (A, T), purine/R (A, G) or pyrimidine/Y (C, T), and amino/M (A, C) or keto/K (G, T). Vertices labelled S, R, and M indicate the percent of nucleotides in a genome characterised as Strong, puRine and aMino respectively. Like the t-SNE plots, labelled virus genomes are coloured and marked according to ICTV Order whereas unlabelled ones are depicted as open grey circles.

\subsection{Nucleotide-based features}

\begin{figure}[!ht]
\centering
\includegraphics[trim={13.5cm 0 0 0},clip,width=4in]{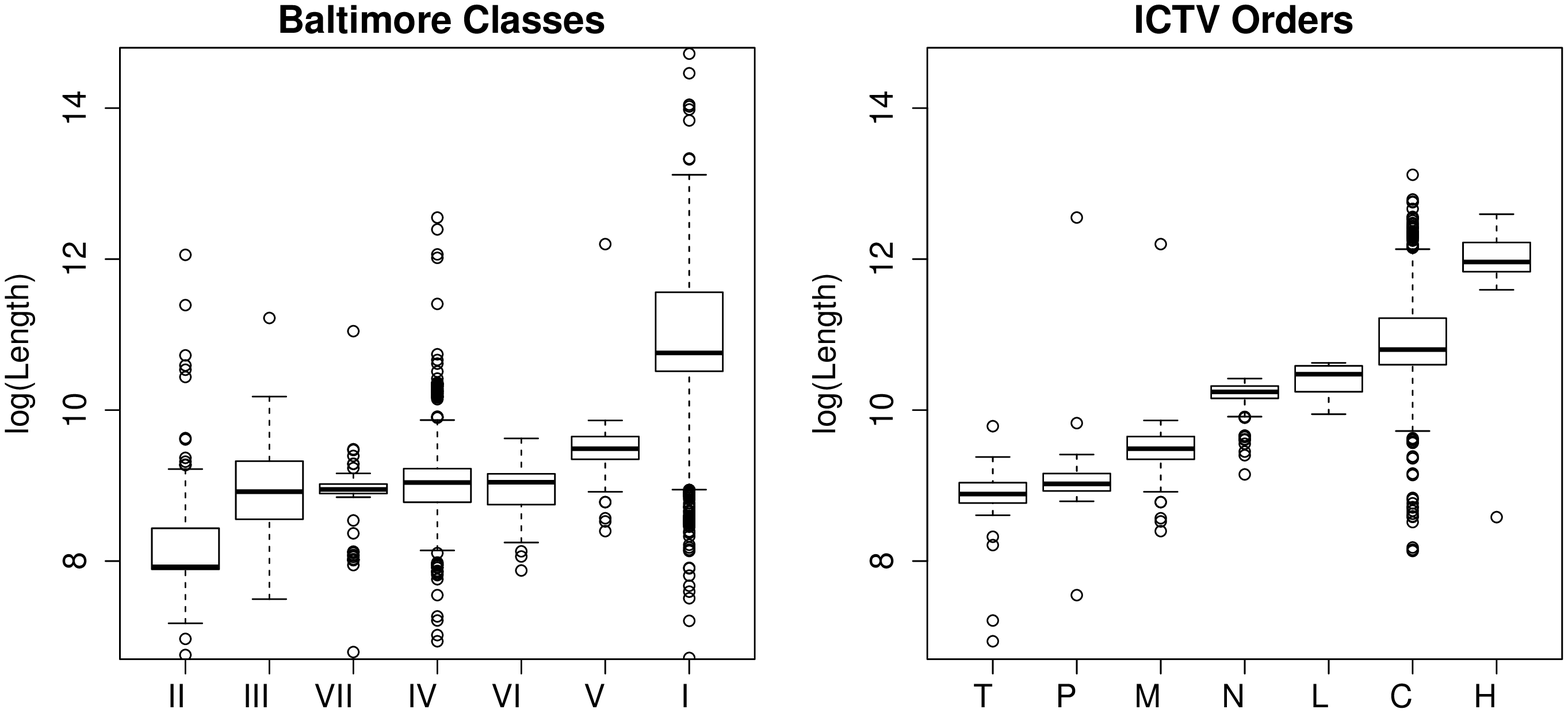}
\caption{Box plot of log(Length) for virus genome sequences assigned to each ICTV Order.}
\label{logNV1}
\end{figure}

Figure~\ref{logNV1} shows the distribution of the simplest descriptive statistic of a sequence $-$ Length $-$ for sequences assigned to a taxonomic class; the seven box plots are arranged by increasing median Length. ICTV Orders can be distinguished fairly well using Length.

\begin{figure}[!ht]
\centering
\includegraphics[width=6in]{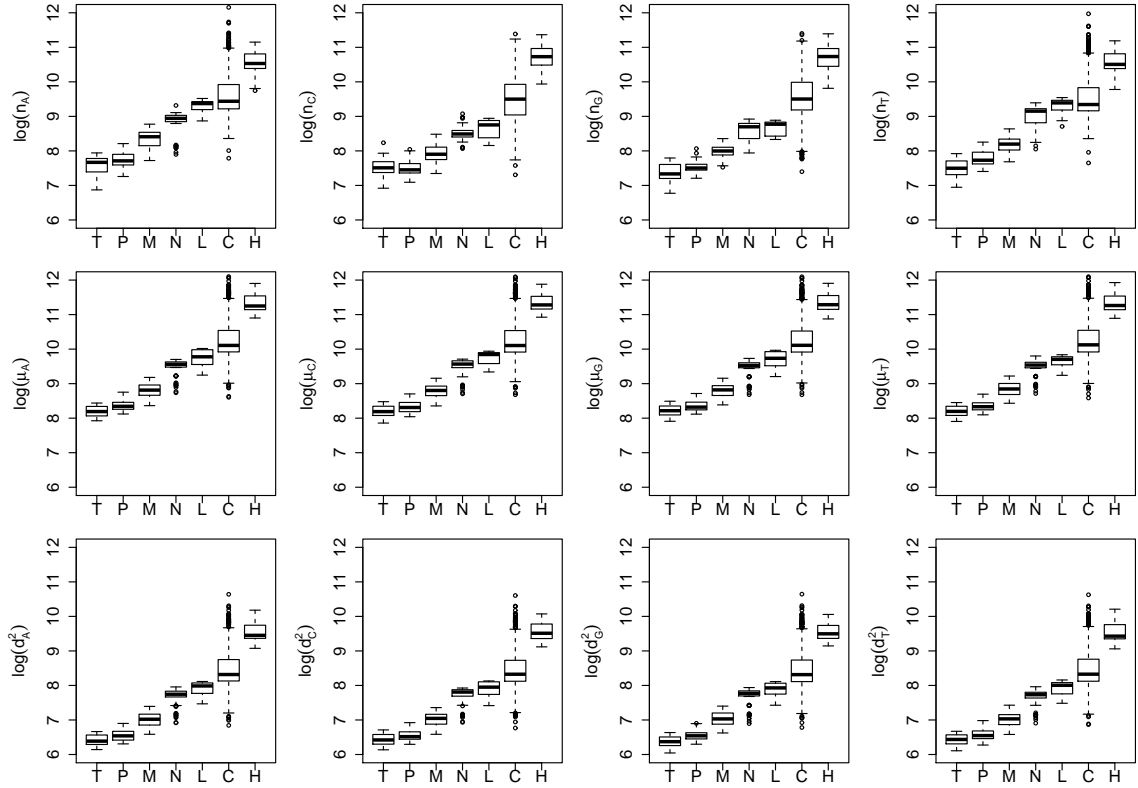}
\caption{Box plots of composition- and location-based statistics for virus genome sequences assigned to each ICTV Order.}
\label{logNV_order}
\end{figure}

Figure~\ref{logNV_order} shows the distributions of slightly more sophisticated summary statistics of a sequence $-$ nucleotide count ($n_A, n_C, n_G, n_T$), mean position ($\mu_A, \mu_C, \mu_G, \mu_T$) and normalised variance of the position ($d^2_A, d^2_C, d^2_G, d^2_T$) $-$ for sequences assigned to each taxonomic class. ICTV Orders can be distinguished reasonably well using every one of these features.

\begin{figure}[!ht]
\centering
\includegraphics[width=6in]{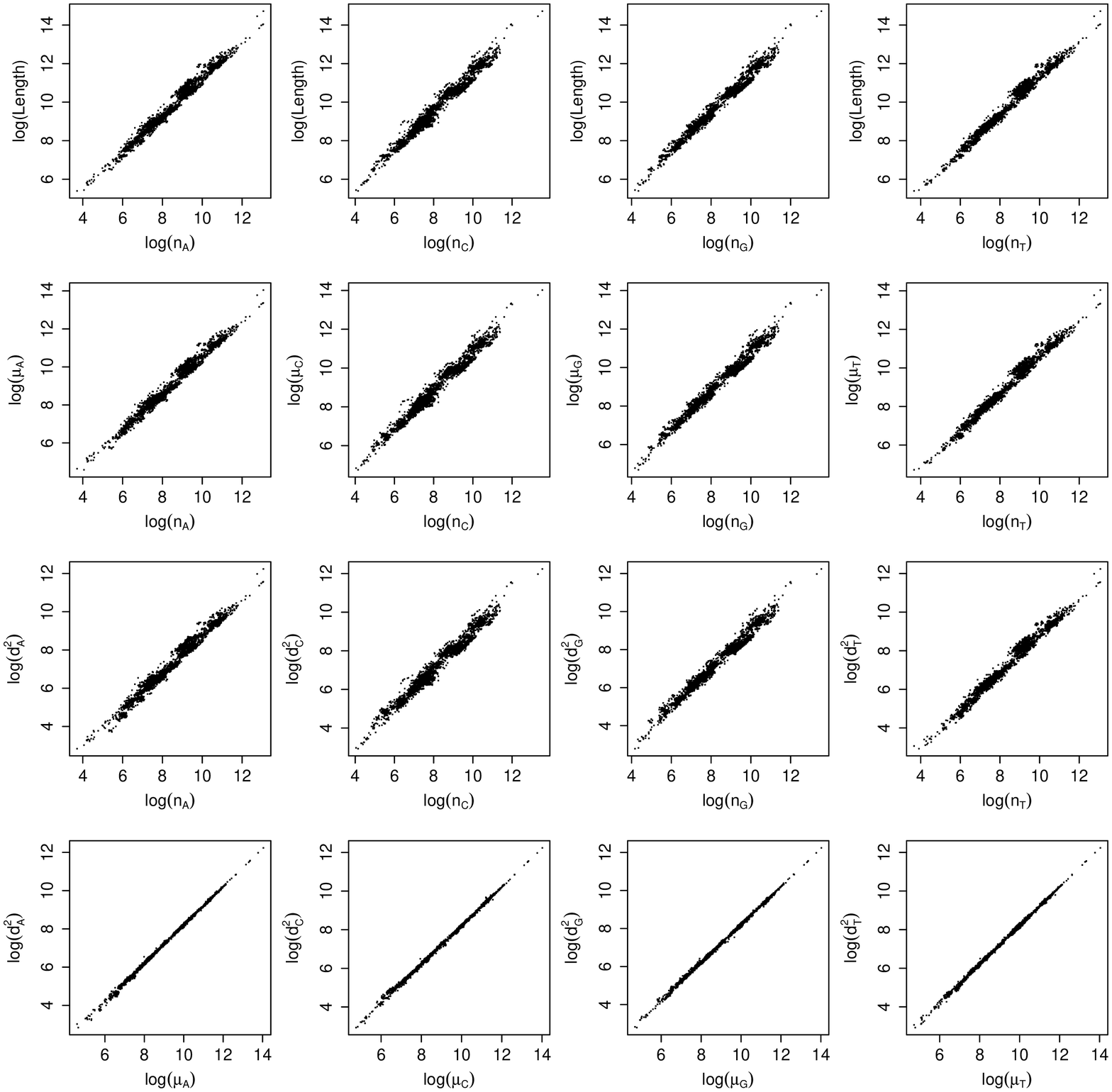}
\caption{Scatter plots between different summary statistics of virus genome sequences.}
\label{varScatterPlot}
\end{figure}

Figure~\ref{varScatterPlot} shows scatter plots for various pairings of the descriptive statistics genome length, nucleotide count, mean position in the sequence and normalised variance of the position. Every plot exhibits a strong linear correlation between the two variables with that for mean position and normalised variance being the most marked. 

\begin{figure}[!htbp]
\centering
  \begin{subfigure}[h]{0.035\textwidth}
    \includegraphics[trim={0 0 20cm 0},clip,height=0.6\paperheight]{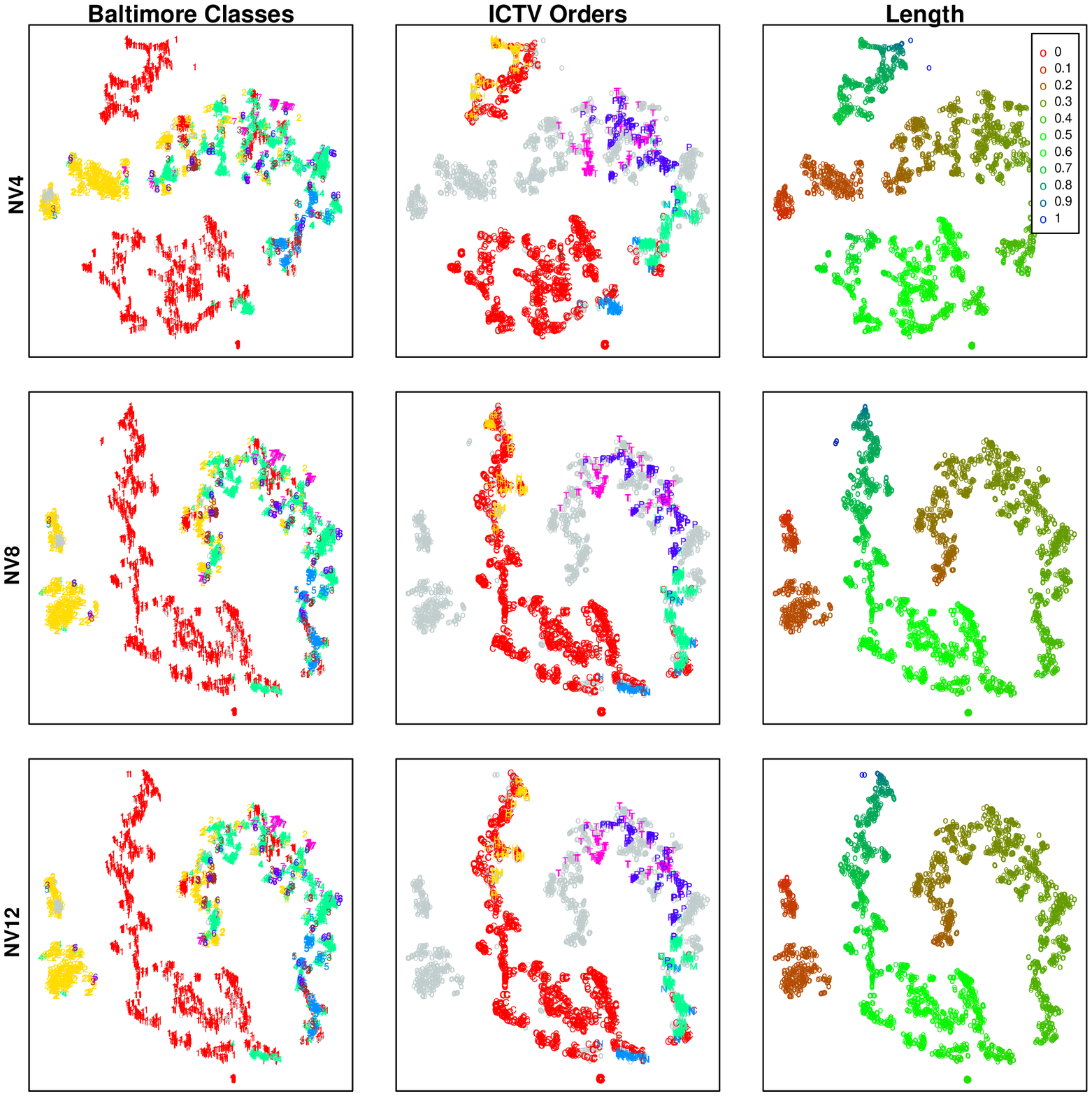}
    \end{subfigure}
    \hspace{-0.5em}%
    \begin{subfigure}[h]{0.786\textwidth}
        \includegraphics[trim={7.5cm 0 0 0},clip,height=0.6\paperheight]{tsne_class_nonSatellites_2.eps}
    \end{subfigure}
\caption{t-SNE plots of the NV4, NV8 and NV12 representations of labelled and unlabelled virus genome sequences with points coloured by ICTV Order or genome length.}
\label{tsne_class}
\end{figure}

Figure~\ref{tsne_class} shows scatter plots of the entire dataset represented using the NV and its derivatives (rows) embedded into two-dimensions where points are coloured by taxonomic class (left column) or sequence length (right column). Viruses assigned the same ICTV Order cluster together, different classes are fairly well separated, and classes with similar Length distributions (Figure~\ref{logNV1}) are neighbours. There is a smooth transition from the shortest sequence (red end of the spectrum) sequence to the longest sequence (blue) $-$ subject to a discontinuity occurring when embedding high dimensional objects into a low dimensional space (see \cite{MaatenJMLR2008} for technical details). The importance of Length in separating different ICTV Orders is also reflected in the results of our subsequent classification experiments.

\subsection{Word-based features}

\begin{figure}[!ht]
  \begin{center}
    \includegraphics[trim={12cm 0 0 0.5cm},clip,width=4in]{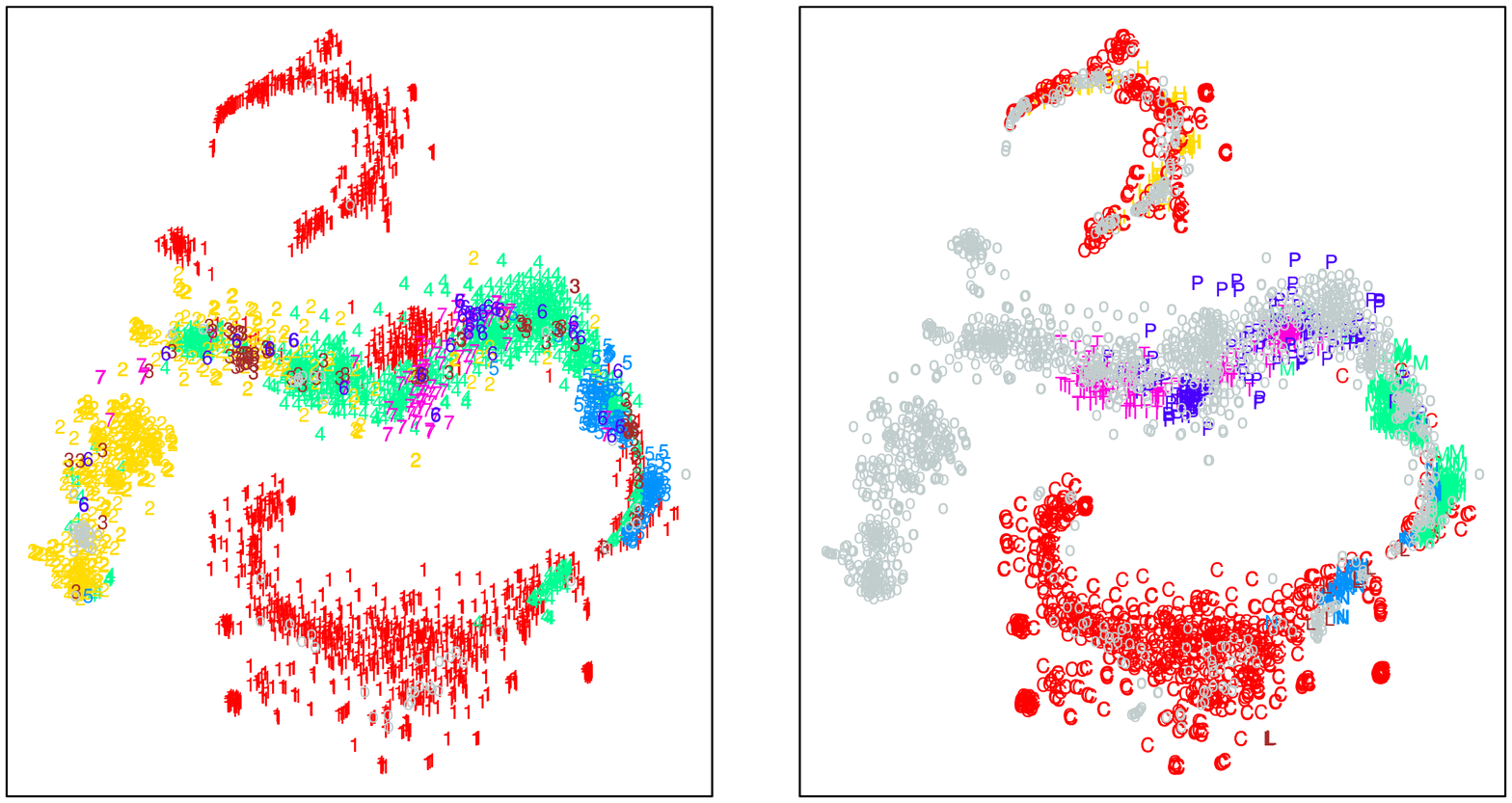}
  \end{center}
  \caption{t-SNE plot of the 6-mer NV representation of labelled and unlabelled virus genome sequences with points coloured by ICTV Order.} 
\label{tsne_class_kmer} 
\end{figure}

Figure~\ref{tsne_class_kmer} shows a scatter plot of the entire dataset represented using the 6-mer NV embedded into two-dimensions where points are coloured by taxonomic class. The cleaner boundaries between ICTV Orders with less overlap between them reflect the fact that the $k$-mer NV is a much richer representation of a sequence than NV12 (Table~\ref{features_kmer}).

\subsection{Compression-based features}

\begin{figure}[!htbp]
\centering
  \begin{subfigure}[h]{4.5in}
  \begin{flushright}
    \includegraphics[trim={0 0 0 10cm},clip,width=5in]{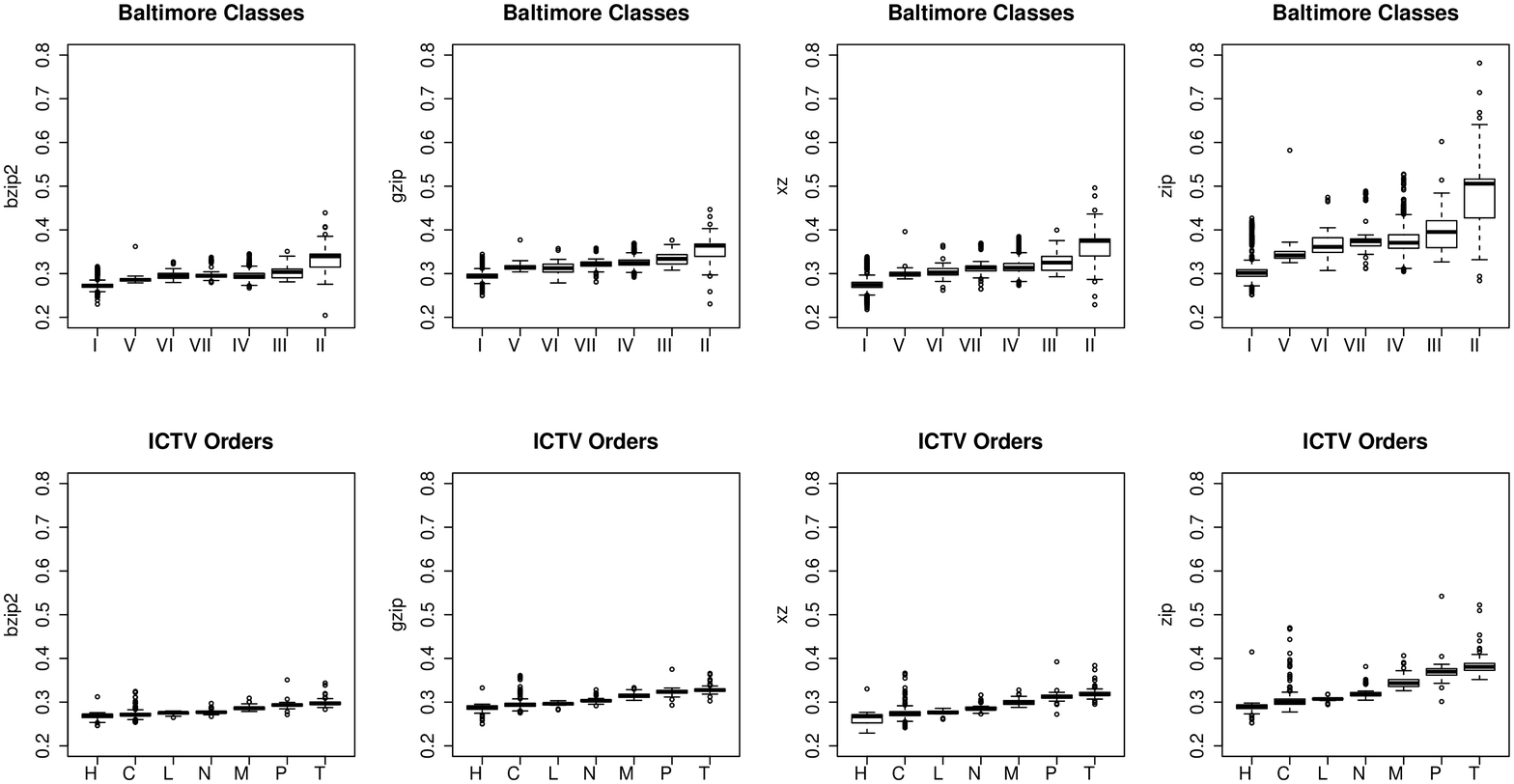}
    \end{flushright}
    \end{subfigure}
    \hspace{-0.5em}%
    \begin{subfigure}[h]{4.5in}
    \begin{flushleft}
        \includegraphics[trim={0 0 0 10cm},clip,width=5in]{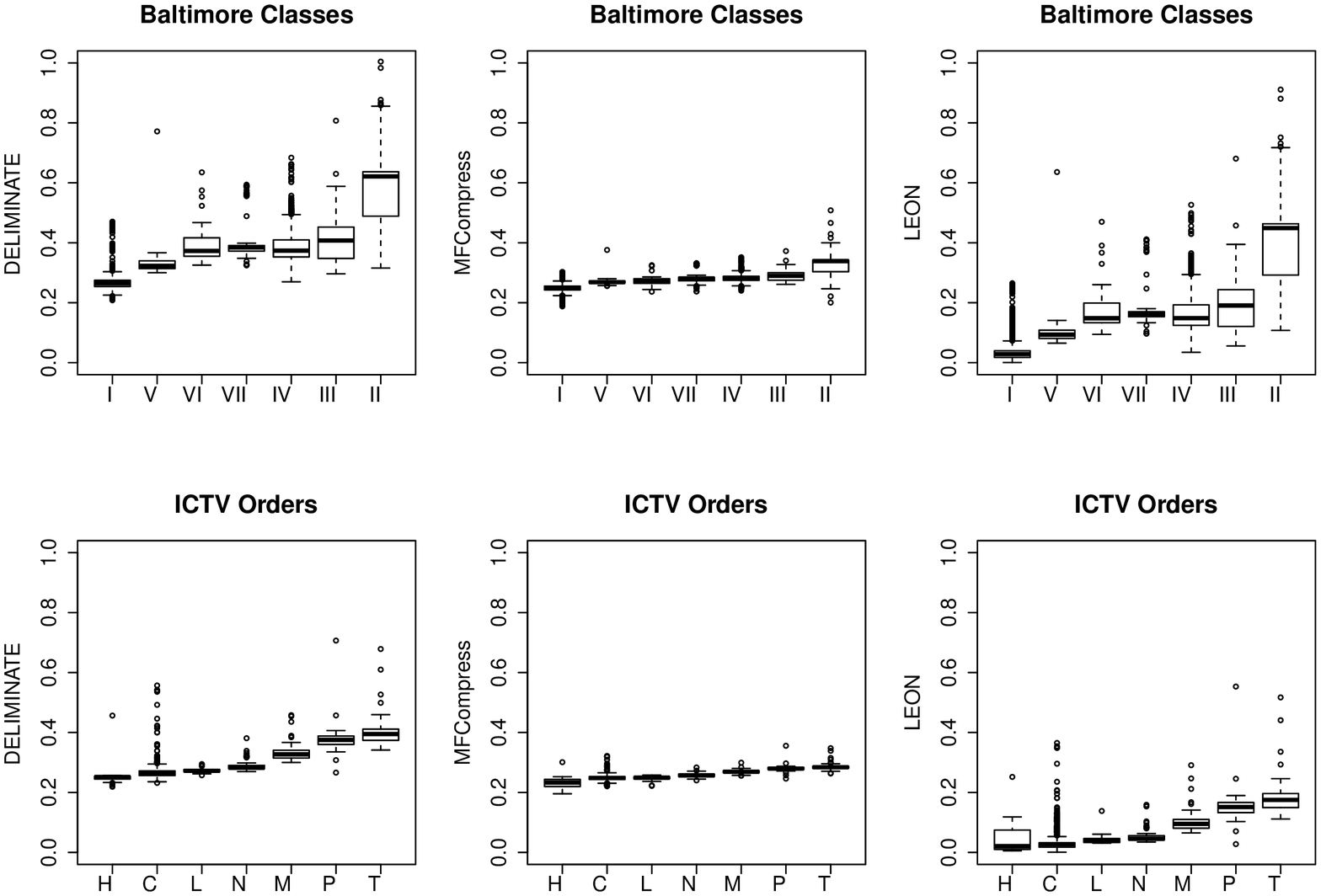}
    \end{flushleft}
    \end{subfigure}
\caption{Box plots of compression ratios achieved by general-purpose (top) and DNA-specific (bottom) compression tools for virus genome sequences assigned to each ICTV Order.}
\label{compRatio_boxplot}
\end{figure}

Figure~\ref{compRatio_boxplot} shows the distributions of different compression ratios of a sequence for sequences assigned to each taxonomic class. Although compression ratios computed using DNA specific tools are somewhat better at separating ICTV Orders than those obtained using general-purpose tools, all are consistently worse than nucleotide statistics (Length, NV4, NV8 and NV12) (Figure~\ref{logNV1} - \ref{logNV_order}). Of the general-purpose tools, bzip2 achieves the lowest average compression ratio across the entire dataset (0.290) but this is only slightly better than the others (gzip: 0.315, xz: 0.303, zip: 0.359). The DNA specific tools are all better with LEON being the best (0.142, DELIMINATE: 0.283, MFCompress: 0.273).

\begin{figure}[!htbp]
\centering
  \begin{subfigure}[h]{4.5in}
  \begin{flushright}
    \includegraphics[trim={0 0 0 9.5cm},clip,width=5in]{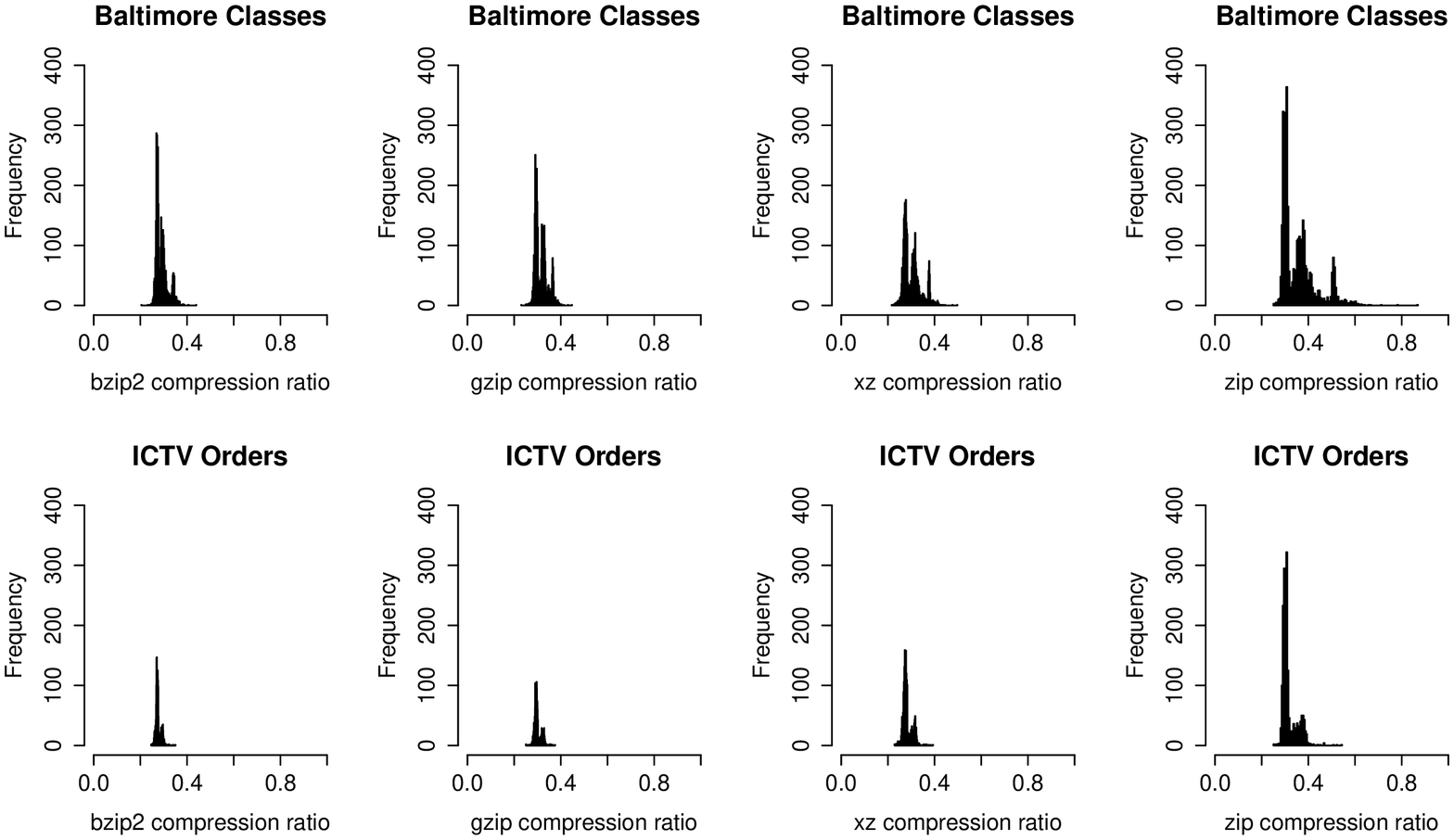}
    \end{flushright}
    \end{subfigure}
    \hspace{-0.5em}%
    \begin{subfigure}[h]{4.5in}
    \begin{flushleft}
        \includegraphics[trim={0 0 0 10cm},clip,width=5in]{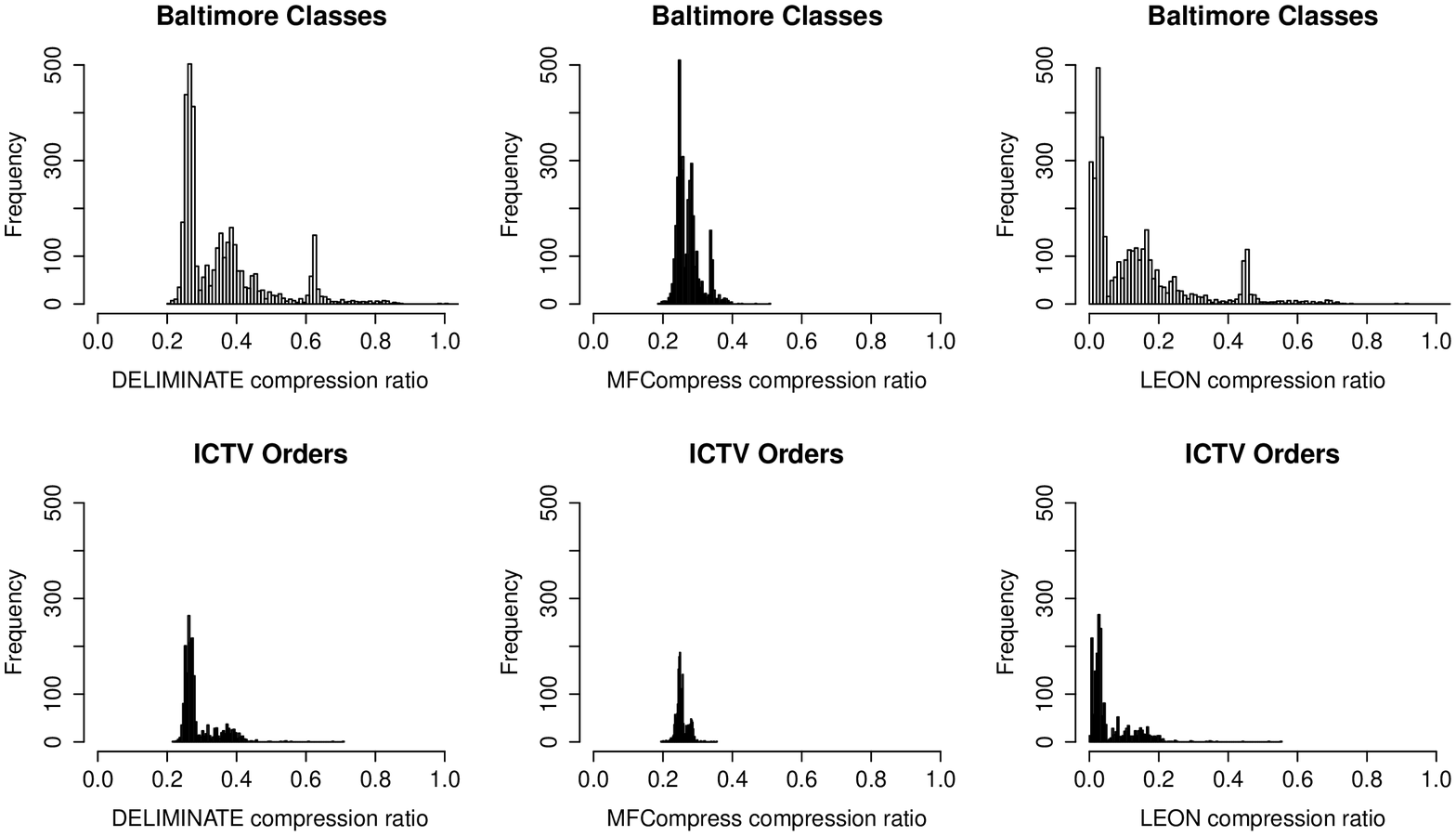}
    \end{flushleft}
    \end{subfigure}
\caption{Histograms of compression ratio achieved by general-purpose (top) and DNA-specific (bottom) compression tools for all virus genome sequences.}
\label{compRatio_hist}
\end{figure}

Figure~\ref{compRatio_hist} shows histograms of compression ratios for the entire dataset. The distributions tend to contain one major peak and two relatively minor ones and skew towards the left side; the degree of skewness is greater for DNA specific tools.

\subsection{Sequence composition based on 2-letter alphabets: SW, RY, and MK}

\begin{figure}[!ht]
\begin{center}
   \includegraphics[trim={0 0 0 11cm},clip,width=5in]{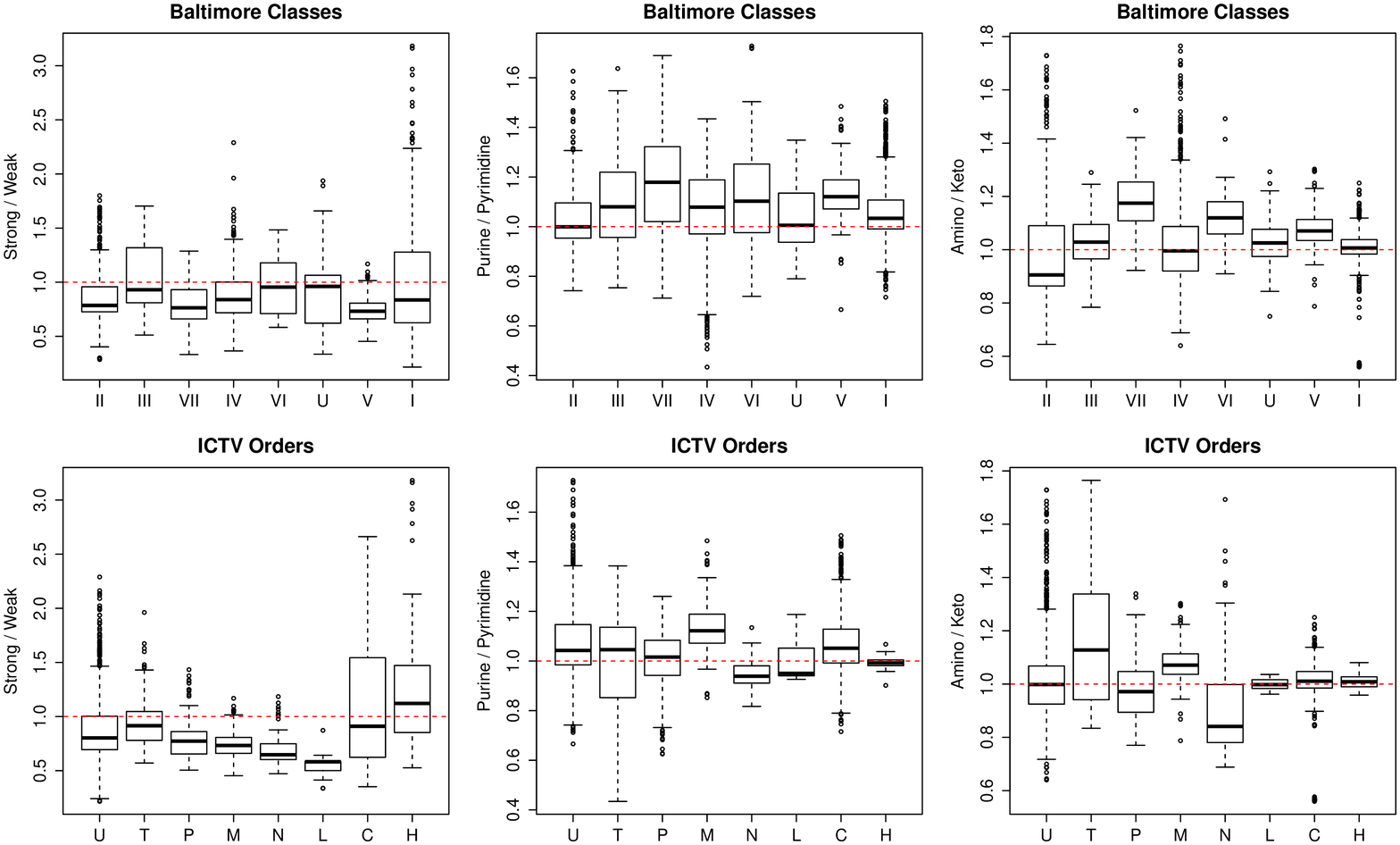}
\end{center}
\caption{Box plots of the reduced alphabet-based nucleotide composition ICTV Order-labelled and unlabelled virus genome sequences.}
\label{compboxplots}
\end{figure}

Figure~\ref{compboxplots} shows the distributions of the ratio of characters in each alternative alphabet $-$ Strong/S (G+C) to Weak/W (A+T), Purine/R (A+G) to Pyrimidine/Y (C+T) and Amino/M (A+C) to Keto/K (G+T) (Table~\ref{features_kmer}) $-$ for sequences with and without (''U'') a taxonomic class assignment; the eight box plots are arranged by increasing median Length and the horizontal red dashed line indicates a ratio of 1.0. The median values of these three ratios are useful guides for grouping and distinguishing between different subsets of ICTV Orders. For example, S/W $>$ 1 for H (the reverse is true for the other 6 Orders), S/W $<$ 1 for C and L, S/W $<$ 1 and M/K $<$ 1 for N and P, and S/W $<$ 1, R/Y $>$ 1 and M/K $>$ 1 for M. Orders with the smallest and largest median values are as follows: S/W (L: min, H: max), R/Y (N: min, M: max), and M/K (N: min, T: max).

\begin{figure}[!ht]
\begin{center}
   \includegraphics[trim={0 0 0 11.5cm},clip,width=6in]{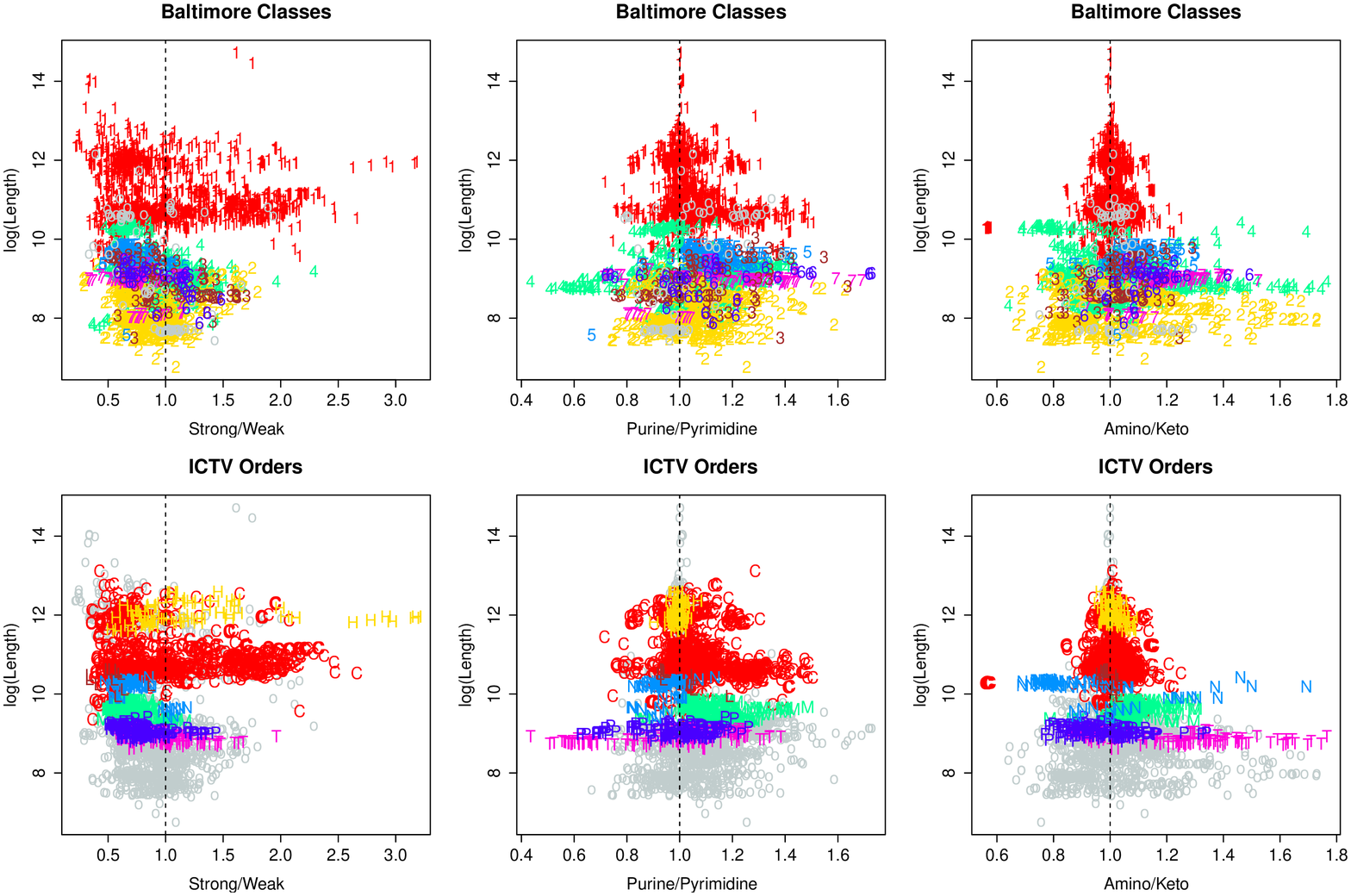}
\end{center}
\caption{Scatterplots of the reduced alphabet-based nucleotide composition of ICTV Order-labelled and unlabelled virus genome sequences and genome length.}
\label{comp}
\end{figure}

Figure~\ref{comp} shows a scatterplot of the natural log transformed value of Length and the ratio of Strong/Weak, Purine/Pyrimidine and Amino/Keto nucleobases in a sequence. Length stratifies the overall reduced alphabet-based composition of virus genome sequences by taxonomic class. Although the ratios are concentrated mostly around one, the degree of dispersion on both sides varies between ICTV Order and alphabet. For example, the pattern of S/W for ICTV Orders H and C and M/K for N and T extend signficantly towards a large ratio, whereas the pattern of M/K for C are roughly symmetric about one. R/Y is the most symmetric and S/W is the least. Our findings are consistent with previous works that focus on specific virus categories \cite{AuewarakulVR2005} and codon usage \cite{BelalovPO2013,VanJGV2016}.

\begin{figure}[!ht]
\begin{center}
   \includegraphics[trim={32cm 3.3cm 0 4cm},clip,width=5in]{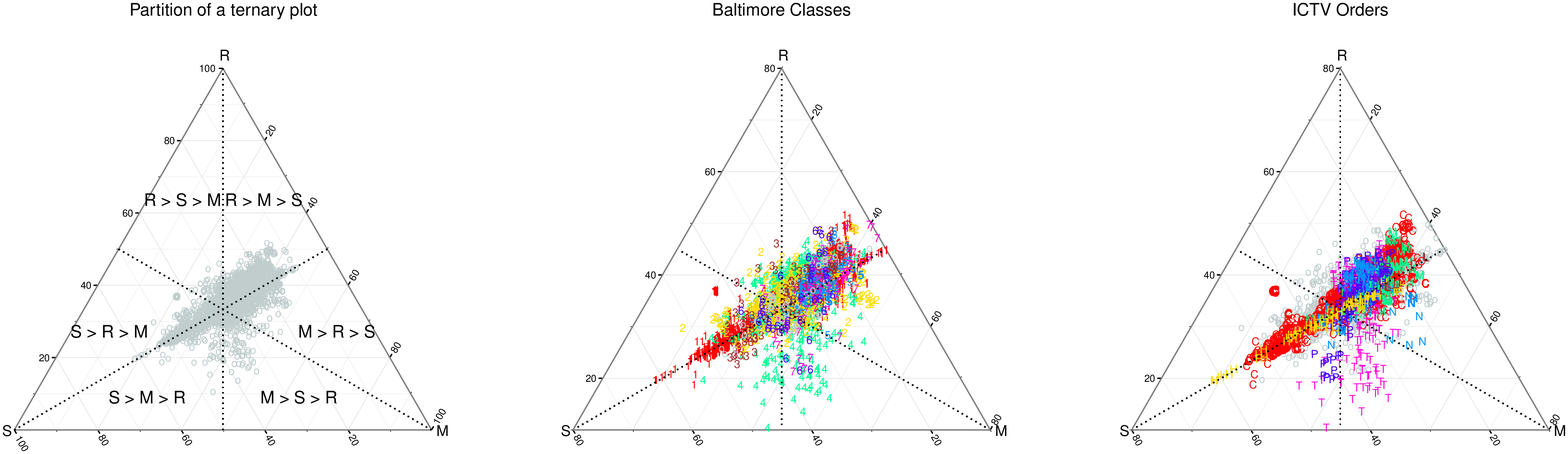}
\end{center}
\caption{Ternary plot of the reduced alphabet nucleotide content of ICTV Order-labelled and unlabelled virus genome sequences .}
\label{ggtern_class}
\end{figure}

Figure~\ref{ggtern_class} displays the relationship between the overall Strong, Purine and Amino content of a sequence using ternary coordinates. Points are concentrated mainly along the median line connecting vertex S to the edge between vertices R and M indicating that S has the largest variance whilst R and M are similar. Different classes tend to favour different ratios and display different patterns. For example, the region M $>$ S $>$ R is dominated by ICTV Order T. ICTV Order H is concentrated at the median line and spans a large range, whereas M and N favour smaller S.

\section{Classification experiments}
\label{sec_results}

\subsection{Methods}
We conducted three different types of experiments.  Common to all three experiments was the following preprocessing and parameter selection procedure.
To improve statistical stability and performance, we normalised features so that each variable in the feature vector has zero mean and unit variance. To tune the parameter(s) of a classifier, we employed 5-fold Cross-Validation over a range of values (Table \ref{paramRange}) in a grid search fashion. The combination of values giving the lowest mean error rate on a set of feature vectors was selected. 
\begin{table}[!ht]
\caption{Range of parameter values used for cross-validation.}
\label{paramRange}
\centering
\begin{tabular}{l l} \hline
\bf Classifier & \bf Range of parameter values \\ \hline
k-NN & $k \in \{1, 2, 3, \cdots, 10\}$ \\ 
RBF-SVM & $C \in \{2^5, 2^6, 2^7, \cdots, 2^{17}\}$ \\
 & $\gamma \in \{2^{-18}, 2^{-16}, 2^{-14}, \cdots, 2^{13}\}$ \\ \hline
\end{tabular}
\end{table}

First, to assess the classification performance of a given combination of feature vectors and classifiers, the 1,865 labelled viruses in the dataset were divided at random into training and testing sets in the ratio $75\%:25\%$. The training set was used to tune the parameters of the classifier using 5-fold cross validation and the classifier with optimised parameter values is used to predict the ICTV Order of the testing set. The result is an error rate for this particular split of the dataset. This procedure is repeated 10 times with random seeds, each giving a different training/testing set division. We report the mean and standard deviation of the testing error rates over 10 different $75\%:25\%$ partitions of the dataset. We plot the confusion matrix of the classification results in order to examine the performance of individual classes. The results are reported in sections~\ref{sec_nuc}, \ref{sec_word} and \ref{sec_best}.

Second, to identify viruses that are difficult to classify, we conducted a leave-one-out experiment where the training set contains all the labelled viruses except for the one to be tested, and the testing set contains one single virus. The experiment is repeated until every labelled viruses has appeared in the testing set. The virus is considered to be difficult to classify if the predicted label does not match the true label. The results are reported in section~\ref{sec_hardVirus}.

Third, to predict the taxonomic class of currently unlabelled viruses, we identified the combination of feature vector and classifier with the smallest mean test errors. We constructed a training set using all 1,865 labelled viruses and used it to learn a classification model. The learned model was used to predict the ICTV Order of every one of the 1,834 unlabelled viruses. The results are reported in section~\ref{sec_unlabelled}.

\subsection{Nucleotide-based features}
\label{sec_nuc}

\subsubsection{Overall classification performance}

Table~\ref{classfErr} shows the mean and standard deviation of the error rate for every combination of classifier and feature vector representation in the task of predicting the ICTV Order of a virus. For a given feature vector, the classifier achieving the lowest mean error rate is highlighted in bold. 

For both classifiers, the error rate decreases from Length to NV12. Length performs respectably, NV4 is a significant improvement over Length but NV8 and NV12 are only marginally better than NV4 $-$ the benefit of adding more complex features (mean position and normalised variance) to nucleotide count is limited. Although k-NN is better with respect to Length, NV4 and NV8, the best classification performance is achieved by the combination of RBF-SVM and NV12.

\begin{table}[!ht]
\renewcommand{\arraystretch}{1.0}
\caption{Classification performance of different nucleotide-based feature vectors and classifiers.}
\label{classfErr}
\centering
\begin{tabular}{l l l} \hline
Feature & \multicolumn{2}{c}{Classifier}\\ \cline{2-3}
vector & k-NN & RBF-SVM \\ \hline
Length & \bf 0.137 $\pm$ 0.013 & 0.144 $\pm$ 0.012 \\ 
NV4 & \bf 0.059 $\pm$ 0.010 & 0.073 $\pm$ 0.009 \\ 
NV8 & \bf 0.051 $\pm$ 0.011 & 0.053 $\pm$ 0.009 \\ 
NV12 & 0.049 $\pm$ 0.007 & \bf 0.038 $\pm$ 0.008 \\ \hline
\end{tabular}
\end{table}

\subsubsection{Simple features can give respectable classification performance}

The EDA supports the classification results (Table~\ref{classfErr}). The plots of Length (Figure~\ref{logNV1} and Figure~\ref{comp}) as well as the composition- and location-based statistics underlying the NV and its derivatives (Figure~\ref{logNV_order}) show these features are fairly good at separating ICTV Orders. The  improvements of NV8 or NV12 over NV4 is explained by the scatter plots between Length, counts, mean position and normalised variance (Figure~\ref{varScatterPlot}): all pairs exhibit a strong linear correlation with that between mean position and normalised variance being the strongest.

The t-SNE plots (Figure~\ref{tsne_class}) show that even in an unsupervised setting, distinct clusters corresponding roughly to ICTV Orders are present. Length plays an important role in the embedding procedure. Classes having similar Length distributions (Figure~\ref{logNV1}) are neighbours in the t-SNE plots, with a smooth transition from the shortest to the longest sequence. The similarity between the NV8 and NV12 embeddings indicates that normalised variance adds a negligible amount of information to mean position, most likely because these features are strongly correlated (Figure~\ref{varScatterPlot}). Thus, total length conveys a large amount of information about a virus genome sequence and its decomposition into the count of each nucleobase underlies most of the discriminatory power of NV12.


\subsubsection{Small classes tend to be confounded with large ones with similar Length distribution}

\begin{figure}[!ht]
\centering
\includegraphics[trim={0 0 9.5cm 0},clip,width=4in]{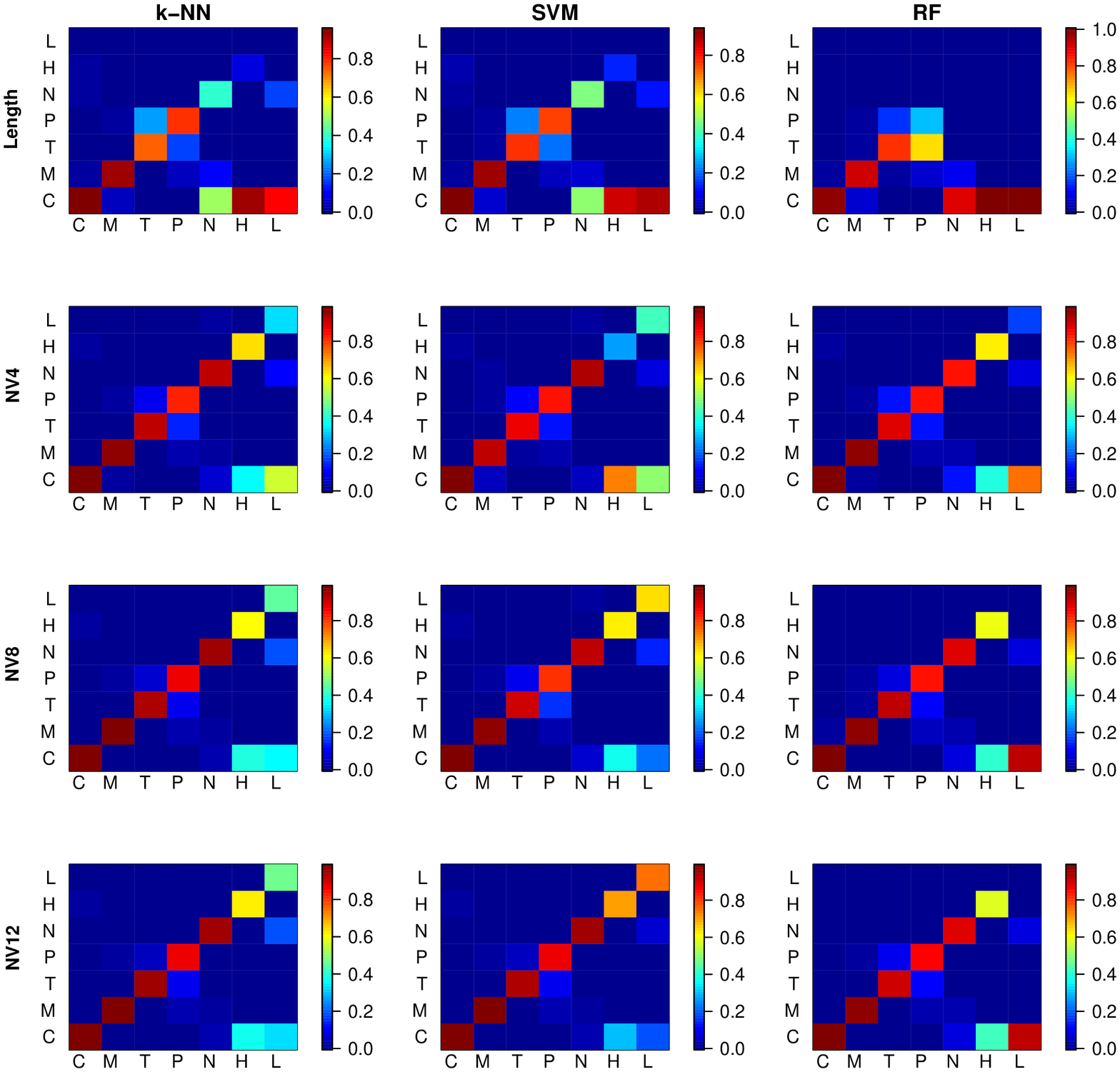}
\caption{Confusion matrices for the classification results of different combinations of nucleotide-based feature vectors and classifiers.}
\label{confMatOrder}
\end{figure}

Figure~\ref{confMatOrder} breaks down the classification errors (Table~\ref{classfErr}) by taxonomic class. In each subplot, the x-axis and y-axis represents the true and predicted ICTV Order respectively. Colours represent the percentage of viruses from one class (column) being predicted as a member of another class (row). Diagonal entries represent rates of correct classification (true class equals predicted class) whilst off-diagonal entries are rates of misclassification. From left to right and bottom to top, classes are ordered according to decreasing size (Table~\ref{virusInSchemes}).

The size (number of members) and error rate of a class are clearly correlated. When using Length with k-NN or RBF-SVM, the accuracy of the larger ICTV Orders C, M, T, P are consistently above 0.7 whereas those of the smaller ICTV Orders N, H, L are generally below 0.5. However, similarly sized classes can have significantly different errors. For example, ICTV Orders H and N have 66 and 67 members respectively but H has a much higher chance of being misclassified. This may be caused by the Length distribution (Figure~\ref{logNV1}) since viruses from H have a higher chance of being misclassified as C.

Class size and Length distribution are key contributors to classification performance. Errors arise mainly from small classes, which tend to be confounded by larger ones with similar Length distribution. However, viruses from large classes are unlikely to be misclassified into small classes with similar Length distribution because all the confusion matrices are non-symmetric (the bottom right has higher values than the upper left). 

In summary, larger classes have higher accuracy and higher false positive rates whilst smaller ones have lower accuracy and higher false negative rates. These results echo the proximity of different classes in the t-SNE plots (Figure~\ref{tsne_class}): large classes such as the ICTV Order C are well separated from the other classes but the ICTV Order H and L are largely confounded by C (indicating higher errors in the confusion matrices).

\subsubsection{Performance improves from NV12 more for small classes than larger classes}

When comparing the confusion matrices from Length to NV12 (Figure~\ref{confMatOrder}), the values along the diagonals increase (an increase in correct classification) and values in off-diagonal regions decrease (a decrease in misclassification). Like overall classification errors, the improvement is noticeable going from Length to NV4 and modest thereafter. However, the behaviour of small classes (for example, ICTV Orders N, H, and L) and large classes (for example, ICTV Orders C, M, T, and P) differ. For small classes, the change is dramatic from Length to NV4, and still noticeable from NV4 to NV12, especially when using the RBF-SVM. For large classes, the improvement from Length to NV4 exists but is much less noticeable than for small classes, and almost negligible from NV4 to NV12. This suggests that a more complex feature vector improves the performance for small classes in particular, but for large ones only marginally so. Owing to the bias in size, the extra performance obtained from small classes seems negligible in terms of the overall classification errors (Table~\ref{classfErr}).

\subsection{Word- and compression-based features}
\label{sec_word}

This section extends the previous section on features based on single nucleotides to features derived from $k$-mers and the entire sequence. The aim is to investigate the predictive power of different features and classifiers and identify the best combination for the task of virus taxonomy.

\subsubsection{Overall classification performance}

Table~\ref{classfErr_summary} shows the best performance achieved for feature vectors constructed from various word- and compression-based features. The best combinations of feature vector and classifier are listed and that with lowest mean error rate is highlighted in bold. 

Richer alphabets improve classification performance. The $k$-mer counts of the 4-letter alphabet (``Counts of ACGT'') performs better than that of the concatenated three 2-letter alphabets (``Concat-counts of SW, RY, MK'') which are better than that of the individual 2-letter alphabets (``Counts of SW,'' ``Counts of RY,'' and ``Counts of MK'') which in turn are better than the 1-letter alphabet (Length).

When ACGT is the alphabet, $k$-mer counts outperform RTD, possibly because RTD has lost crucial information on genome length. Although the performance of a feature vector constructed by concatenating $k$-mer counts and RTD (``Concat-counts and RTD of ACGT'') is better than RTD, it is still worse than $k$-mer counts.

The $k$-mer NV of ACGT significantly outperforms NV12 (1-mer NV) (Table~\ref{classfErr}). Whilst longer $k$-mers contain richer information, they do not always improve performance because the 6-mer feature vector is not the best. For $k$-mer NV of ACGT, as $k$ increases, error rates follow the same trend in that they first decrease and then increase. The cause may be redundancy and noise. Likewise concatenating statistics of subwords of a $k$-mer (``Concat-$k$-mer NV of ACGT'') does not improve performance over $k$-mer NV; $k$-mer counts of ACGT outperform $k$-mer NV of ACGT. 

By analysing SVM weights of ``Concat-$k$-mer NV of ACGT,'' we find that counts of the largest $k$-mers are associated with higher weights. This agrees with our findings in nucleotide-based features that counts play a key part in classification while mean position and variance are far less important. These results are encouraging because the dimension of the feature vector can be reduced by more than a third from the concatenated $k$-mer NV to the simpler $k$-mer counts.

Compression-based feature vectors underperform $k$-mer based feature vectors as their features are relatively coarse, but their error rates are still respectable. Compression ratios derived from DNA-specific tools outperform those from general-purpose ones.

\begin{table}[!ht]
\renewcommand{\arraystretch}{1}
\caption{Classification performance of different word- and compression-based features and classifiers.}
\label{classfErr_summary}
\centering
\begin{tabular}{l l l} \hline
Features           & Best combination & Error rate \\ \hline
Length & Length, k-NN & 0.137 $\pm$ 0.013 \\ 
Counts of SW & 6-mer, k-NN & 0.052 $\pm$ 0.012 \\ 
Counts of RY & 6-mer, RBF-SVM & 0.030 $\pm$ 0.006 \\ 
Counts of MK & 6-mer, RBF-SVM & 0.044 $\pm$ 0.008 \\ 
Concat-counts of SW, RY, MK & 6-mer, RBF-SVM & 0.016 $\pm$ 0.006 \\ 
Counts of ACGT & 4-mer, RBF-SVM & \bf 0.006 $\pm$ 0.002 \\ 
RTD of ACGT & 3-mer, RBF-SVM & 0.014 $\pm$ 0.004 \\ 
Concat-counts and RTD of ACGT & 4-mer, RBF-SVM & 0.009 $\pm$ 0.004 \\ 
$k$-mer NV of ACGT & 4-mer, RBF-SVM & 0.007 $\pm$ 0.002 \\ 
Concat-$k$-mer NV of ACGT & 4-mer, RBF-SVM & 0.007 $\pm$ 0.003 \\ 
General-purpose compression & CRGP, RBF-SVM & 0.092 $\pm$ 0.007 \\ 
DNA-specific compression & CRDNA, RBF-SVM & 0.089 $\pm$ 0.008 \\ 
Compression-based & CRA, RBF-SVM & 0.073 $\pm$ 0.011 \\ \hline
\end{tabular}
\end{table}

\subsection{Using the best feature vector -- classifier combination to address biological questions}
\label{sec_best}

A feature vector constructed using 4-mer counts of ACGT and the RF-SVM classifier gives the lowest error rate (Table~\ref{classfErr_summary}). This feature-classifier combination is used in our experiments in the following sections.

\subsection{Identifying viruses that are difficult to classify}
\label{sec_hardVirus}

Table~\ref{diffVirus} lists the 8 out of 1,865 analysed viruses that are misclassified (the true label does not match our predicted label) and of these, 6 are from small classes (H, N and L). Table~\ref{classfErr} and Figure~\ref{confMatOrder} provide an explanation in terms of both overall error rates and those of small classes.

\begin{table}[!ht]
\renewcommand{\arraystretch}{1}
\caption{Difficult viruses.}
\label{diffVirus}
\centering
\begin{tabular}{l l l l} \hline
Accession No. & Name & Label & Predicted \\ \hline
NC018874.1 & Abalone herpesvirus  &  H   & C \\ 
	   & Victoria/AUS/2009	& & \\
NC005830.1 & Acidianus filamentous virus 1 & L   & M \\ 
NC009965.1 & Acidianus rod-shaped virus 1 & L   &  C  \\ 
NC024709.1 & Ball python nidovirus & N  &   C  \\ 
NC001609.1 & Enterobacteria phage P4 & C &     T   \\ 
NC002515.1 & Mycoplasma phage P1 & C   &  P \\ 
NC005881.2 & Ostreid herpesvirus 1 & H   &    C \\ 
NC019413.1 & Sulfolobales Mexican & L   &   C  \\ 
		& rudivirus 1 && \\ \hline
\end{tabular}
\end{table}

\subsection{Predicting the ICTV Order of unlabelled viruses}
\label{sec_unlabelled}

We used the best combination of feature vector (4-mer counts of ACGT) and classifier (RBF-SVM) to train a model from all labelled virus genome sequences. We employed the learned model to predict the ICTV Order of all unlabelled virus genome sequences.
Table~\ref{predClass_kmer} gives the number of currently unlabelled viruses predicted in each taxonomic class.

\begin{table}[!ht]
\caption{Predicted ICTV Order for unlabelled viruses.}
\label{predClass_kmer}
\centering
\begin{tabular}{l l l l l l l} \hline
C & H  & L & M  & N   & P     & T \\ \hline
 273 & 80 & 0 & 85 & 12 & 403 & 981 \\ \hline
\end{tabular}
\end{table}

\section{Discussion}
\label{sec_discussion}

\subsection{Overview of this work}

In this work, we performed a systematic study aimed at comparing and contrasting the predictive power of various features of virus genome sequences and classifiers in a virus taxonomy task. We find that for nucleotide-based features, the simple Length already gives respectable performance, with a mean error rate of 0.137. NV4 gives a significant improvement but including higher order moments reduces overall error rates only marginally. Given the importance of Length, smaller classes tend to be confounded with larger classes that have similar Length distributions; performance on these classes can be improved with more sophisticated features and classifiers.

We find that a richer alphabet improves performance, as well as a suitable $k$-mer, but including higher order moments or using longer $k$-mers is not always beneficial. Compression-based features perform worse than most $k$-mer features, with those based on tools tailored for DNA sequences outperforming those based on general-purpose ones. The best feature-classifier combination in our study is 4-mer counts and RBF-SVM (the lowest mean error rate of 0.006). Using the best experimental settings determined in the study, we identified 8 out of 1,865 viruses as difficult to classify. These may warrant further attention from virologists (for example, it is possible that the ``true'' label present in the RefSeq file may be incorrect). We also predicted the ICTV Order for unlabelled sequences.

\subsection{Validation of ICTV Order predictions using an updated dataset}

The NCBI RefSeq dataset is continuously updated as new genomes are sequenced, taxonomy criteria are updated and scientific discoveries are made. A number of changes were made to the dataset since we started the work. A version downloaded on 20$^{th}$ August 2017 contains 3,721 non-satellite non-segmented viruses, an increase of 22 compared to the dataset used in this paper. This more recent version of the dataset contained 101 new viruses but 79 viruses in our version had been removed. We found that 17 out of 1,834 viruses that had no ICTV Order in our earlier version of the dataset had an ICTV Order assignment in the later version: 16 from {\em Caudovirales} and 1 from {\em Tymovirales}. Of these, we had correctly predicted 16 and misclassified one. The taxonomic labels for the misclassified virus ({\em Xaxnthomonas phage Cf1c}) had changed significantly: the later version of the dataset had assigned it to a completely different ICTV Family and Genus.

The changes to the later dataset are small compared to the one we used for the study so the classification models and findings remain relevant. However, the experimental framework can handle significant changes in the dataset such as the addition of two new Orders by the ICTV.

\subsection{Imbalanced classes}

The virus genome sequence dataset is imbalanced, with larger classes having significantly more samples than smaller ones. Small classes consistently give higher error rates, a general issue associated with almost any classifier as their aim is to minimise the overall error rate. From a statistical perspective, a small number of samples may not be representative of their population, hence models built on them are less able to generalise to unseen samples. 

Efforts were made to improve performance for small classes. We tried resampling methods \cite{BekkarIJDKP2013,HeTKDE2009} but found that although they can be beneficial to small classes, they are harmful for large ones and degrade overall classification performance. However, we found that more sophisticated features and classifiers are useful especially for small classes; in particular, we observe noticeable improvements when the features are changed from Length to NV12, then from NV12 to $k$-mer features.

Nonetheless, insufficient data is an inherent disadvantage of small classes that is unlikely to be eliminated completely. The same problem also holds for human experts, who are more likely to misclassify unfamiliar or rare viruses. 

A practical consequence of applying classifiers trained using a biased data set is that label predictions for a small class imply a high level of confidence in the assignment, whereas predictions of a large class do not. Hence, for the currently unlabelled viruses, we would be more confident in the predicted class membership of viruses if they are classified into a small class such as ICTV Order N, H or~L.

\section{Future work}
\label{sec_future}

\subsection{Word-based feature vectors}

We studied features based on the $k$-mer statistics where a precise match was required between a sequence of $k$ consecutive nucleotides and a given $k$-mer. A natural extension is to allow mismatches with gapped $k$-mers \cite{LodhiJMLR2002, LeimeisterB2017} which are short sequences similar to the given $k$-mers but containing different patterns or having different lengths. 

\subsection{Hierarchical classification}
The current classification is performed only at the Order level. A natural extension is to identify the entire taxonomic path of a virus by classifying from Order to Species. 
The challenge of classification at deeper levels is that the number of classes is large and the number of samples per class is small. 
Hierarchical classification is a potential solution as it exploits the hierarchical relationship between classes and thus improves classification performance;
however, while there are examples of its application to various biological sequences such as gene expression and protein sequences~\cite{SillaDMKD2011,FreitasRTDMTA2007}, there is no explicit study demonstrating its use in the classification of viruses into ICTV taxa based on reference genome sequences. Related works, including \cite{YuPO2013,HernandezACUL2013,HuangJTB2016} typically use flat classification for individual ICTV taxonomic levels without accounting for hierarchy.

\subsection{Distributed representations of complete viral genome sequences: word and sequence embeddings}

In our current work, the feature variables were all manually designed. In the future, we can explore the efficacy of feature representations generated by automated methods. For instance, the word2vec methodology \cite{MikolovACUL2013}, originally proposed for NLP, generates vector representations for words using neural network language models. A similar model has been successfully applied to construct features for biological sequences \cite{AsgariPO2015}.

In this work, the features used to represent complete viral genome sequences were all hand-crafted. In the future, we can explore the efficacy of features produced using automated methods for a variety of virus taxonomy-, phylogeny-, and genome biology-related tasks. For instance, the word2vec methodology trains a (shallow) Neural Network over a large text corpus in order to generate a mapping of words or phrases in the lexicon to vectors of real numbers in a low-dimensional space such that words with similar linguistic context are close by in the Euclidean space. Originally proposed for Natural Language Processing, the word2vec framework has been used to construct features for biological (BioVec) and protein (ProtVec) sequences \cite{AsgariPO2015}. seq2vec, an extension of word2vec that estimates a representation of a complete protein sequence as opposed to $k$-mers, outperforms ProtVec on protein classification problems \cite{KimothiACUL2016}. Embedded representations of proteins have been used for protein property prediction tasks \cite{yang2018learned}. For sequence retrieval and classification tasks, a Mahalanobis distance metric in the embedded feature vector space may be better than the default Euclidean metric \cite{kimothi2017metric}.

\subsection{Graph-based semi-supervised learning}

The dataset we analysed contains a large proportion of unlabelled virus genomes (Table \ref{virusInSchemes}). Because we frame virus classification as a supervised learning (SL) problem, these samples were discarded when training the k-NN and SVM models. Semi-supervised learning (SSL) \cite{Zhu2009} uses unlabelled data in conjunction with a small amount of labelled data to yield models whose classification performance can surpass that which could be obtained either by discarding the unlabeled data (SL) or by neglecting the labels (unsupervised learning). Graph-based SSL \cite{Zhu2009} (GSSL) uses a graph reflecting some notion of closeness between samples to infer the class membership of unlabelled samples. In this way, GSSL may be seen to propagate the limited number of initial labels present in the training data through the graph. SSL has been applied to different biological sequences and used for tasks such as protein function prediction \cite{MetzUKWCI2009} and gene function prediction \cite{SantosESA2014}. To date, it has not been applied to virus genome sequences. We suspect the S3VM \cite{BennettNIPS1999}, an SSL version of the SVM based on self-training, will be better at predicting ICTV Order than our SVM with the lowest error rate.

We propose that GSSL provides a way to generate new insights about virions from a graph whose topology is informed by a large number of sequence relationships and whose discrete- and/or real-value node labels are informed by sparser information on virus biology. Specifically, consider the following network of objects and their relationships: a node denotes a virus, an edge specifies the phylogenetic distance between the genomes of the connected viruses (for example, a divergence metric between low-dimensional vectors constructed from nucleotide-, word-, compression- or embedding-based features), and a node label indicates a phenotypic property such as taxon (for example, ICTV Order or Baltimore Class), host (eukaryote, archaeon or bacterium), and capsid type (for example, helical or icosahedral). A GSSL-based framework makes it easy to incorporate, integrate and exploit diverse incoming data: the flood of genomic sequences adds new nodes to the graph whilst the trickle of biological properties adds to the small but growing pool of node labels. Together, the two permit inferences to be made about viruses for which there is no or limited knowledge apart from their complete genome sequences. Furthermore, the graph's community structure provides the basis for learning a data-driven taxonomy of the virosphere where taxa are organised not as a tree but as a network whose structure (the number of nodes and the pattern of edges between individual as well as groups of related viruses) can evolve in light of new data.

\noindent \textbf{Acknowledgements}: The authors gratefully acknowledge Dr K. Bryson and Dr D. Fernandez-Reyes from University College London for discussions.

\bibliographystyle{plain}
\bibliography{VirusGenomeRef}

\end{document}